\documentclass[letterpaper,twocolumn,10pt]{article}
\usepackage{usenix}

\usepackage{cite}
\usepackage{amsmath,amssymb,amsfonts}
\usepackage{graphicx}
\usepackage{subcaption}
\usepackage{algorithmic}
\usepackage{textcomp}
\usepackage{xcolor}
\usepackage{xspace}
\usepackage{booktabs}
\usepackage{csquotes}
\usepackage{tabularx}
\usepackage{array}
\usepackage{comment}
\usepackage{enumitem}
\usepackage{mdframed}
\usepackage{xurl}

\setlist[itemize]{noitemsep,topsep=0pt}

\newcommand{\mypara}[1]{%
    \vspace{5pt}\noindent\textbf{#1}%
}

\newcommand{\tn}[1]{%
    {\textcolor{magenta}{Tushar: #1}\xspace}%
}
\newcommand{\rs}[1]{%
    {\textcolor{red}{RS: #1}\xspace}%
}

\newmdenv[
    backgroundcolor=gray!5,
    linecolor=gray,
    outerlinewidth=1pt,
    roundcorner=5pt,
    skipabove=2pt,
    skipbelow=0pt,
    innerleftmargin=3pt,
    innerrightmargin=3pt,
    innerbottommargin=3pt,
    innertopmargin=3pt,
    font=\normalsize
]{shadedboxed}

\begin{document}

\title{%
    \Large\bfseries
    Understanding Users' Security, Privacy, and Trust
    Concerns in Healthcare AI Systems
}


\author{
\fontsize{10.5}{11}\selectfont
\begin{tabular}[t]{@{}c@{\hspace{3.5em}}c@{\hspace{3.5em}}c@{}}
{\rm Tushar Nayan} & {\rm Mohammed Al Duniawi} & {\rm Song Fang} \\
Florida Intl.\ University & Florida Intl.\ University & Florida Intl.\ University \\
tnaya002@fiu.edu & maldu010@fiu.edu & sfang@fiu.edu \\[6pt]
{\rm Aokun Chen} & {\rm Arjun Arunasalam} & {\rm Ruimin Sun} \\
Florida State University & Florida Intl.\ University & Florida Intl.\ University \\
ac25cv@fsu.edu & aarunasa@fiu.edu & rsun@fiu.edu
\end{tabular}
}


\date{\today}
\maketitle

\begin{abstract}

Artificial intelligence (AI) is increasingly integrated into healthcare systems, from clinical information retrieval to workflow automation and decision support, which introduces new security, privacy, and trust (SP\&T) risks with direct implications for patient care. 
While prior work has examined these risks from technical and regulatory perspectives, less is known about how these risks are naturally discussed in online communities.
To address this gap, we conduct a large-scale analysis of 14 medical communities on Reddit and extract 6,514 relevant posts. Through systematic qualitative coding and LLM-assisted classification, we investigate concerns expressed in healthcare AI discussions, their prevalence, and how they evolve over time.
Our analysis reveals that users perceive healthcare AI as expanding traditional healthcare trust boundaries and attack surfaces through integrations with electronic health records, third-party services, cloud platforms, and automated clinical workflows. 
Posts express concerns about over-privileged access to sensitive records, opaque data flows, unauthorized secondary use, and unsafe automation can amplify the consequences of errors and misuse. 
%
%
%
%
Beyond concern expression, we identify four attitudes toward healthcare AI risks: exploratory efforts to understand these risks, such as asking who is accountable for AI-related clinical errors, cautious acceptance conditional on stronger safeguards, including human oversight, transparency, regulation, and technical protections, while others remain skeptical about system trustworthiness or become resigned to perceive inevitability to increasing data exposure and automation. 
Longitudinal analysis further shows that SP\&T discussions have grown significantly over time, with notable shifts coinciding with major healthcare AI deployments, data privacy incidents, human oversight, and regulatory developments. 
Our findings characterize emerging attack surfaces, shifting trust relationships, and opaque data-governance practices discussed in the corpus, providing actionable guidance for system design, deployment, and governance.

Together, our findings characterize the prevalence of healthcare AI concerns, variations in expressed attitudes, and evolving perceptions of risk, informing the design and governance of safer, more transparent, and accountable healthcare AI systems.

\end{abstract}

\section{Introduction}

Artificial intelligence (AI) is increasingly integrated into healthcare to improve clinical decision-making, diagnostic accuracy, operational efficiency, and access to care~\cite{topol2019high, esteva2019guide}. Recent advances in large language models (LLMs) ~\cite{chatgpt, claude, gemini, singhal2025toward, google2025medgemma, chen2023meditron} and AI agents~\cite{shimgekarAgenticAIFramework2025, caetanoAgenticWorkflowsConversational2025, openai2026healthcare} have enabled the automation of healthcare workflows involving reasoning, tool use, and interactions with multiple information systems~\cite{collaco2026role}. Healthcare organizations have begun deploying ambient clinical agents such as Dragon Copilot, which  generate clinical documentation, referral letters, after-visit summaries, and electronic health record (EHR) updates from patient encounters~\cite{microsoft2025dragoncopilot}. Beyond documentation, AI systems are increasingly used for care coordination, patient communication, and administrative workflows~\cite{patel2025priorauthagents, commure2026platform}. 

As emerging AI systems become integrated into healthcare, unique security, privacy and trust (SP\&T) risks arise.
For instance, LLM-based agents are vulnerable to prompt injection, retrieval and memory poisoning and tool misuses attacks that can manipulate agent behavior or trigger unauthorized actions~\cite{greshake2023more,liu2023prompt, willison2023promptinjection,zhang2025agent}. At the same time, healthcare AI raises privacy and trust concerns, including sensitive attribute inference, training data reconstruction, lack of explainability and hallucinated clinical content, among other documented concerns~\cite{lehman2021does, carliniExtractingTrainingData2021, singhal2023large, tu2024towards, mesko2023imperative,collaco2026role}

While prior work has identified SP\&T risks from technical and regulatory perspectives, how these risks are discussed in healthcare-related online communities remains poorly understood~\cite{greshake2023more, collaco2026role, liu2023prompt, singhal2023large}. Prior user-centered studies rely on surveys, interviews, and evaluations on specific systems such as AI-assisted triage, clinical decision support, and ChatGPT-based consultation tools~\cite{steerling2025influences, kelly2025investigating, benda2026qualitative, du2024exploring}. 
Consequently, these studies provide limited visibility into how healthcare AI risks are discussed in everyday online settings.~\cite{collaco2026role, liu2023prompt}. 
As healthcare organizations increasingly deploy AI agents across clinical workflows, understanding these perceptions remains an important open challenge.

To address this gap, we conduct a large-scale analysis of online discussions to investigate SP\&T concerns and responses expressed in healthcare AI discussions. Specifically, we answer the following research questions: 
\textbf{RQ1}: What SP\&T risks do users perceive and discuss regarding healthcare AI, particularly emerging agentic systems? 
\textbf{RQ2}: How do attitudes toward healthcare AI vary across perceived SP\&T risks and user groups? 
\textbf{RQ3}: How have discussions of healthcare AI risks evolved over time as AI technologies become increasingly integrated into healthcare workflows?

To answer these questions, we collect and analyze a dataset of approximately 5.8 million posts from the top 14 healthcare-related communities on Reddit. We combine manual annotation, supervised classification, qualitative coding, and LLM-assisted analysis to identify discussions related to SP\&T concerns surrounding healthcare AI. This process yields $\sim$6.5K SP\&T-related posts, which we use to develop a taxonomy of perceived risks, characterize user attitudes toward healthcare AI, and examine how these discussions evolve over time. We further apply interrupted time series analysis to investigate the impact of major healthcare AI deployments and technological milestones on public discourse surrounding SP\&T risks \footnote{Codebook, event timeline, 
and supporting materials: \url{https://github.com/tusharnayan10/medAI}.}


Our mixed-methods analysis reveals that users' concerns are dominated by issues of clinical oversight, safety, and accountability rather than traditional privacy risks. The most frequently discussed concerns involve human oversight of healthcare AI (31.6\%) and clinical safety (29.4\%), reflecting concerns about the consequences of AI errors, overreliance on automation, and the need for clinicians to remain responsible for patient care. Posts also frequently discuss healthcare-specific model training (13.1\%)  and the explainability of AI decisions (8.6\%), highlighting concerns about how medical AI systems are developed, validated, and integrated into clinical workflows. In contrast, discussions of personally identifiable information (1.7\%) and third-party data sharing (0.5\%) are comparatively rare. Instead, privacy concerns are more commonly framed around the collection and use of activity/behavioral data (5.5\%) and the governance of healthcare AI through compliance and regulatory safeguards~(5.7\%).

We also identify diverse expressed attitudes toward healthcare AI across the analyzed discussions, ranging from reactive efforts to protect privacy and security to information-seeking, skepticism towards healthcare organizations/AI vendors and resignation towards AI's growing role in care. Our longitudinal analysis further shows that these perceptions evolve alongside the healthcare AI ecosystem, with discussions of SP\&T risks fluctuating in response to new deployment, regulatory actions and other relevant incidents. 
%

Overall, our findings suggest that trust in healthcare AI depends not only on protecting sensitive data but also ensuring that AI systems operate safely, transparently, and accountably within real-world healthcare environments.We conclude by deriving implications for AI vendors, healthcare organizations, and regulators.

This paper makes the following contributions: 
\begin{itemize}
    
    \item \textbf{Large-scale empirical study of healthcare AI risk perceptions.} We present a large-scale analysis to characterize SP\&T concerns expressed in online healthcare AI discussions. 
    \item \textbf{Characterization of expressed concerns and attitudes.} Through qualitative and quantitative analysis, we identify the dominant SP\&T concerns associated with healthcare AI and characterize variation in expressed attitudes toward these concerns.     
    \item \textbf{Longitudinal analysis of healthcare AI discourse.} We examine the changes in expressed concerns around major healthcare AI developments, providing design, deployment, and governance insights for healthcare AI. 
\end{itemize}

\section{Background and Related Work}

\subsection{Background}

AI supports every stage of the healthcare workflow by analyzing health-related data to generate summaries, recommendations, and other clinical content~\cite{rajpurkar2022ai,moor2023foundation}. Applications include detecting cancer~\cite{mckinney2020international}, predicting outcomes from electronic health records (EHRs)~\cite{jiang2023health}, answering medical questions~\cite{singhal2023large}, and summarizing notes from recorded clinical encounters~\cite{lukac2025ambient}. These uses make security, privacy, and trustworthiness central considerations.


\mypara{Security Concerns.} Within healthcare, security concerns involve protecting AI-supported healthcare workflows from unauthorized access and manipulation. Prior work has examined adversarial attacks, data poisoning, model manipulation, unauthorized access, and compromised healthcare data pipelines, which may affect clinical decisions and healthcare operations~\cite{finlayson2019adversarial, ma2021understanding}. More recently, LLM-integrated applications and AI assistants have been shown to be susceptible to prompt-injection attacks and retrieval-based manipulations that can  influence downstream medical decisions~\cite{greshakeNotWhatYouve2023,greshake2023more}. 

\mypara{Privacy Concerns.} Privacy concerns involve the collection, use, sharing, and retention of sensitive health information. These risks arise from model memorization, unintended disclosure of sensitive medical information, and secondary use of healthcare data beyond its original purpose~\cite{lehman2021does, el2015anonymising}. Privacy-preserving approaches, including federated learning, differential privacy, and secure computation, seek to reduce data exposure during model development and use~\cite{rieke2020future,kaissis2020secure,li2019privacy}.

\mypara{Trust Concerns.} Trust concerns encompass AI safety, explainability, fairness, and human oversight as mechanisms for supporting responsible deployment~\cite{tjoa2020survey, esmaeilzadeh2021patients, amann2020explainability, lee2021included,  shevtsova2024trust, kauttonen2025trust, cutillo2020machine}. Recent studies further highlight safety concerns associated with medical hallucinations and clinically unsupported recommendations generated by LLM-based systems, which may undermine patient trust and introduce risks to clinical decision-making~\cite{singhal2023large, ji2023survey, jahan2025}.

While prior work has examined these risks from technical assessments and deployment perspectives, less is known about how online users discuss healthcare AI, including which SP\&T concerns they raise, how they interpret these risks, and what attitudes they express toward them.

\subsection{Related Work}

\mypara{User Studies of Healthcare AI Concerns.}
Recent work has examined patients and clinicians perceptions of healthcare AI through surveys, interviews, focus groups, scenario-based studies, and evaluations of particular applications~\cite{esmaeilzadeh2021patients,lee2021included,shevtsova2024trust,buschMultinationalAttitudesAI2025, heinrichs2025physicians}. Richardson et al.~\cite{Richardson2021}, for example, conducted 15 focus groups and identified patient are concerned about safety, choice, data integrity, and security, while clinicians~\cite{Abdelwanis} highlight training, workflow integration, privacy, and accountability. While these approaches provide valuable insights, they primarily capture perceptions elicited under researcher-defined settings. Consequently, little is known about how SP\&T concerns are naturally discussed among users who care about healthcare AI.

\mypara{Online Discussions on SP\&T in General. } Online discussion platforms (e.g., Reddit, X, and Stack Overflow) provide a valuable lens into expressed perceptions of and responses to emerging technologies in real-world settings.
Unlike surveys and interviews, which elicit responses to researcher-defined questions, online discussions capture organically expressed concerns, experiences, misconceptions, and coping strategies.
Prior research has leveraged Reddit and other social media platforms to investigate users' privacy concerns, security behaviors, and trust perceptions toward technologies such as smart-home devices and conversational AI systems~\cite{de2014mental,fiesler2018reddit,liItsConsumerBe2023,aliUnderstandingUsersSecurity2025}. While In healthcare, Tang et al.~\cite{fpubh} combined topic modeling,
sentiment analysis, and temporal analysis of Reddit discussions. These studies demonstrate that online discussions can reveal nuanced concerns and behaviors that may not be readily observable through traditional user-study methodologies.

Our study connects these research strands through a focused analysis of SP\&T concerns and attitudes in healthcare-related Reddit communities. We distinguish concerns from attitudes.  Our contribution therefore lies in examining SP\&T discourse within healthcare contexts, where sensitive patient information, clinical responsibilities, and patient-safety considerations shape the discussion. This analysis complements participant studies with evidence from discussions that arise without researcher prompting and adds an SP\&T-focused perspective to broader analyses of healthcare AI discourse.

\section{Methodology}
\label{sec:methodology}

To investigate SP\&T concerns and attitudes surrounding healthcare AI systems, we conduct a large-scale analysis of Reddit~\cite{reddit} threads, including posts and their associated comments from healthcare-related subreddits.

\begin{figure}[!h]
    \centering
    \includegraphics[width=\columnwidth, trim=0.8cm 0.1cm 0.5cm 0cm, clip]{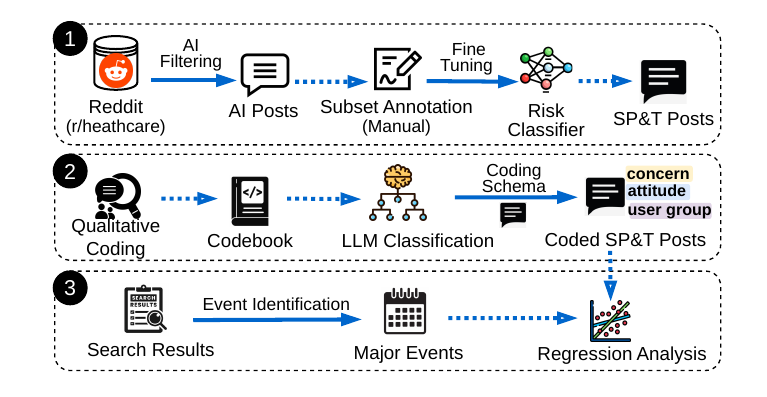}
    \caption{Methodology overview.}
    \label{fig:overview}
\end{figure}

Figure~\ref{fig:overview} presents an overview of our methodology. Our analysis proceeds in three stages. First, we collect healthcare-related Reddit discussions, identify posts related to healthcare AI, and train a classifier to detect SP\&T-related content. Second, we conduct qualitative coding on SP\&T posts to develop a codebook capturing expressed concerns and attitudes, and discussion context. Using this codebook as the labeling schema, we apply an LLM-assisted classification pipeline to scale the annotations across the full set of SP\&T\-related posts for prevalence analysis. Third, we identify major AI and healthcare-AI events from public search results and use regression analysis to examine whether and how SP\&T discussions shift around these events.

\subsection{SP\&T-related Post Identification}
\label{subsec:spt_identification}

Reddit~\cite{reddit} is a large online platform organized into topic-specific communities (subreddits), where Reddit participants post content and engage in threaded discussions.
These organic discussions make Reddit a suitable source for studying expressed perceptions of AI in healthcare contexts~\cite{proferes2021studying}. 
We identify $14$ medical and healthcare-related subreddits (e.g.,  r/AskMedical, r/healthcare) based on their relatively high weekly engagement rates.
These subreddits cover clinical practice, medical education, healthcare work, and patient-facing questions, thereby capturing wider perspectives on AI usage in healthcare. 
Detailed information on subreddit selection (and full list) are provided in Appendix~\ref{subsec:healthcare_subreddits} and Table~\ref{tab:subreddits}.

Using the ArcticShift API~\cite{arcticshift2026}, we collect posts spanning January 2020 to March 2026, covering a period marked by the rapid emergence of foundation models and their increasing integration into healthcare workflows. 
Our raw dataset consists of approximately 5.8 million posts in total, with $\sim$717K submissions, and $\sim$5.11M comments, from $\sim$1.5M distinct Reddit accounts users. 
The scale and longitudinal coverage of this dataset enable us to analyze how SP\&T concerns emerge, evolve, and respond to major developments in the healthcare AI ecosystem. We applied standard preprocessing to clean the data (Appendix~\ref{subsec:data_preprocessing})


\mypara{Keyword Design and Filtering}
\label{subsubsec:keyword_filtering}
Given the scale of the dataset, identifying healthcare AI discussions within a large corpus of medical discussions requires careful keyword design. A naive keyword-based approach often suffers from low coverage and high false positives, particularly in medical subreddits, where participants frequently share medical experiences, symptoms, diagnoses, treatments, and clinical terminology that may not be related to AI. To address this challenge, we adopt an iterative keyword expansion strategy. We begin by compiling an initial set of keywords drawn from prior literature \cite{carliniExtractingTrainingData2021}, \cite{zhang2023sa, gumusel2024user}, \cite{yan2024protecting}, \cite{williamson2024balancing}. This includes general AI terminology as well as medical-specific AI expressions. We then iteratively refine the keyword set by (i) retrieving posts using the current keyword list, (ii) inspecting matched posts to assess relevance, (iii) identifying new terms, synonyms, and variations appearing in posts and comments; (iv) expanding and updating the keyword list accordingly. This iterative process allows us to capture evolving language patterns and reduces the risk of missing relevant discussions. 
Keyword matching was performed over each thread's title, selftext, and top five relevant comments, yielding 11,981 candidate healthcare AI threads. 
The final dictionary contains 208 AI-related keywords. Frequently matched keywords are presented in Table~\ref{tab:keyword_categories} of Appendix~\ref{subsec:keyword_seed_corpus}.

\mypara{}{Manual Annotation}
\label{subsubsec:manual_annotation}
Because manually reviewing $\sim$12K posts is impractical. Therefore,  we created a seed dataset to train an SP\&T classifier. We sampled 1,400 posts (100 from each of 14 subreddits) and manually annotated them for SP\&T-related content. To ensure quality annotation, two security and privacy experts jointly annotated 200 posts to establish annotation guidelines, then independently annotated the remaining posts. Inter-annotator agreement was high, with Cohen's Kappa of $\kappa = 0.94$ and 95.3\% overall agreement. The final dataset contained 584 SP\&T-related posts (41.7\%) and 816 non-SP\&T-related posts.




\mypara{SP\&T Model Training}
\label{subsubsec:model_training}
To identify SP\&T-related discussions at scale, we fine-tune a binary classifier using the manually labeled dataset of 1,400 Reddit posts. The dataset was split using stratified sampling into training, validation, and test sets (70:10:20). We evaluate several transformer-based models, including BERT~\cite{bert}, RoBERTa~\cite{roberta}, DeBERTa~\cite{deberta}, Llama-3~\cite{llama3}, and Phi-4~\cite{phi4}. Among these models, DeBERTa achieved the best held-out performance, with 92\% accuracy and a 91\% F1-score (Table~\ref{tab:model_classification_result}). We therefore select DeBERTa as the SP\&T classifier and apply it to the full corpus of 11,981 posts, identifying 6,514 SP\&T-related discussions. To further validate the classifier beyond the held-out test set, we conduct a secondary manual audit of 150 model-labeled posts sampled from this corpus. The classifier achieved 97.3\% accuracy and a 96.8\% F1-score on this audit set, providing additional evidence of SP\&T identification pipeline reliability. Appendix~\ref{subsec:post-over-time} and Figure~\ref{fig:spt_vs_ai} further report the temporal volume of SP\&T-related discussions relative to all healthcare AI discussions.

\subsection{Data Analysis}
\label{subsec:data_analysis}

\mypara{Sampling and Coding}
\label{subsubsec:sampling_coding}
To identify SP\&T concerns expressed in healthcare AI discussions, we conducted an iterative thematic analysis of posts classified as SP\&T-related. We developed an initial codebook using the keywords and the manual annotation process described in Subsection~\ref{subsec:spt_identification}. Two authors then iteratively sampled and coded SP\&T-related posts over a seven-day period, updating the codebook after each coding round to capture newly emerging concerns and refine ambiguous category boundaries.
We organized recurring concerns into themes and sub-themes.
In particular, one post may express more than one concerns, so we consider a primary theme and optional secondary theme and subthemes.
We continued the coding process until reaching thematic saturation, where additional posts no longer introduced substantially new themes, sub-themes, or coding rules as described in Appendix~\ref{subsec:theme_annotation}. Finally, the codebook was developed and refined using an additional 230 SP\&T-related posts, resulting in six themes and thirteen subthemes as detailed in Table~\ref{tab:annotation_codebook}.
We further labeled the posts with medical application domain and discussion type, which provides context for the SP\&T concerns from each post.


\begin{table}[!h]
\centering
\caption{Distribution of posts across SP\&T themes, subthemes, and contextual categories. Percentages indicate the share of posts in each category.}
\label{tab:annotation_codebook}
\footnotesize
\setlength{\tabcolsep}{2pt}
\renewcommand{\arraystretch}{0.95}
\setlength{\aboverulesep}{1pt}
\setlength{\belowrulesep}{1pt}
\begin{tabular}{@{}ll@{}}
\toprule
\multicolumn{2}{@{}l@{}}{\textbf{Themes and Subthemes (\%)}} \\
\midrule
\multicolumn{2}{@{}l@{}}{\begin{tabular}[t]{@{}l@{}}
Trust and Responsible AI (46.02\%)\\
\quad Human oversight (31.60\%); Safety (29.56\%);\\
\quad Explainability (8.62\%); Bias (3.01\%)
\end{tabular}} \\
\addlinespace[1pt]
\multicolumn{2}{@{}l@{}}{\begin{tabular}[t]{@{}l@{}}
Data Usage (16.01\%)\\
\quad Model training (13.11\%); Third-party sharing (0.40\%)
\end{tabular}} \\
\addlinespace[1pt]
\multicolumn{2}{@{}l@{}}{\begin{tabular}[t]{@{}l@{}}
Data Collection (13.97\%)\\
\quad Activity and behavioral data (5.52\%); PII (1.67\%)
\end{tabular}} \\
\addlinespace[1pt]
\multicolumn{2}{@{}l@{}}{\begin{tabular}[t]{@{}l@{}}
Reg.\ \& Legal Compliance (12.03\%)\\
\quad HIPAA (3.52\%); FDA (1.71\%);\\
\quad EU regulatory compliance (0.48\%)
\end{tabular}} \\
\addlinespace[1pt]
\multicolumn{2}{@{}l@{}}{\begin{tabular}[t]{@{}l@{}}
Security Vulnerabilities (6.95\%)\\
\quad Platform security (0.38\%);\\
\quad AI agents/workflow security (0.24\%)
\end{tabular}} \\
\addlinespace[1pt]
\multicolumn{2}{@{}l@{}}{\begin{tabular}[t]{@{}l@{}}
Data Retention (5.02\%)\\
\quad Data retention (0.17\%)
\end{tabular}} \\
\midrule
\textbf{Medical AI domains (\%)} & \textbf{Discussion types (\%)} \\ \midrule
Clinical AI (48.90\%) & General discussion (30.59\%) \\
Administrative AI (19.74\%) & Tool Experience (29.26\%) \\
Patient-facing AI (18.28\%) & Workflow / operation (10.88\%) \\
Development/research AI (6.73\%) & AI product (10.73\%) \\
Other AI (6.33\%) 
& Other (18.54\%) \\ 
\bottomrule
\end{tabular}
\end{table}


\mypara{Prevalence Analysis}
\label{subsubsec:prevalence_analysis}
To analyze the prevalence and evolution of SP\&T concerns, we use GPT-4o~\cite{openai2024gpt4o} to extend the manually developed taxonomy from Section~\ref{subsubsec:sampling_coding} to the full post corpus. Following prior work on LLM-assisted post-hoc data labeling~\cite{aliUnderstandingUsersSecurity2025}, we provided GPT-4o~\cite{openai2024gpt4o} with each Reddit thread, along with the annotation instructions, codebook definitions, labeling rules, and a few manually annotated few-shot examples. The model assigned structured labels for the SP\&T theme and sub-theme, along with secondary contextual metadata, including  application domain and post discussion type. The concern labels form the basis of findings in Section~\ref{sec:users_spt_concerns} and Section~\ref{sec:longitudinal_analysis}, while the application-domain and discussion-type labels provide context to interpret where and how these concerns arise. 

We evaluate the LLM-assisted annotation on our 230 manually coded SP\&T posts used during codebook development. GPT-4o~\cite{openai2024gpt4o} achieved 95\% accuracy and an F1-score of 0.95 at the theme level, and 94\% accuracy and an F1-score of 0.94 at the sub-theme level. Detailed theme- and sub-theme-level classification results are reported in Appendix~\ref{subsec:llm_annotation_performance}. After validation, we applied the annotation framework to the entire dataset and conducted a secondary manual audit of 150 GPT-4o-labeled posts sampled from the full SP\&T corpus. On this audit set, GPT-4o achieved 97\% accuracy and an F1-score of 0.90 for themes and sub-themes. Table~\ref{tab:annotation_codebook} shows the distribution and prevalence of SP\&T concerns.

\mypara{Longitudinal Analysis}
\label{subsubsec:longitudinal-method}
To examine temporal changes in SP\&T concerns expressed in healthcare AI discussions. We first compiled a timeline of significant healthcare AI-related events. Similar with prior work~\cite{aliUnderstandingUsersSecurity2025}, 
we manually reviewed public events on Google search results related to healthcare AI developments. Event collection focused from 2022 to January 2026. Search terms were derived from our qualitative analysis (Sections~\ref{sec:users_spt_concerns}).
We constructed a timeline of candidate events and verified event details through official announcements and primary sources. We then assessed their public visibility through Google searches, checking for coverage across multiple news outlets and recurrence across search rounds. The final event set included major foundation model releases, healthcare AI deployments, regulatory and governance developments, and security and privacy incidents. In total, we identified 41 major events. 
Rather than constructing a comprehensive timeline, we focused on key events likely to shape user discussions and perceptions of healthcare AI risks

For event-specific temporal analysis, we consolidated related sub-themes from our qualitative coding into four high-level categories: (i) AI Development, (ii) Data Privacy, (iii) Human Oversight and Accountability, and (iv) Regulatory and Security Governance.
%
%
For each category, we conducted interrupted time-series analyses of weekly post frequencies using an eight-week pre- and post-event window. We used Ordinary Least Squares (OLS) regression to model weekly discussion frequencies as a function of: (i) a time variable representing the progression of weeks within the event window, (ii) an intervention term (0 before and 1 after the event) capturing immediate post-event changes, and (iii) a time-intervention interaction term capturing changes in post-event discussion trends. 

Because we estimated multiple event--category models and tested both the immediate-level and post-event trend coefficients, we controlled the false discovery rate using the Benjamini--Hochberg procedure at $q < 0.05$. The correction was applied jointly to the $p$-values for the intervention and time--intervention coefficients across all event--category models. We classified an event--category association as a statistically detectable increase when at least one of these coefficients was positive and its Benjamini--Hochberg-adjusted $p$-value was below 0.05. Complete results, including estimates, confidence intervals, and raw and adjusted $p$-values, are reported in
Appendix~\ref{tab:event_category_results}. Finally, we manually examined discussions surrounding statistically significant events to interpret how user discourse shifted following major healthcare AI developments.


\section{SP\&T Concerns in Healthcare AI Discussions}
\label{sec:users_spt_concerns}


The analyzed discussions raise concerns about healthcare AI model behavior and its interactions with healthcare data, infrastructure, and decision-making processes. We organize these concerns by SP\&T theme and subtheme. Healthcare AI application-domain and discussion-type labels provide supporting context for interpreting where and how these concerns arise.



\subsection{Security Vulnerabilities}
\label{subsec:security-concerns}
The discussions raise security concerns about AI workflows and the healthcare platforms with which
they interact.


\subsubsection{AI Workflow Security}
\label{subsubsec:runtime_operational_security}
These discussions concern risks associated with autonomous AI operation and workflow execution.


\mypara{Prompt Injection and Safety Bypass.}
A recurring concern in the discussions was that adversarial prompts could manipulate healthcare AI, override safeguards, or alter clinical outputs. Discussions surrounding \textit{``LLMs (GPT-5, Gemini 2.5 Pro, Claude 4.5 Sonnet) are highly vulnerable to prompt injection''} describe prompt injection as a way for \textit{``malicious people to hijack the LLM's usual behavior''}. Reddit users specifically highlight that fabricated evidence may be inserted into the model through hidden inputs (e.g., \textit{``a completely white-out text not seen by humans''}).

As a result, LLMs outputting \textit{``contraindicated medical advice, including examples where models were induced to recommend thalidomide in a hypothetical encounter with a pregnant patient''}. 
Same concerns appear in discussions surrounding autonomous prescription renewal systems (e.g., \textit{``Artificial intelligence begins prescribing medications in Utah''}). 
Discussions react strongly with concerns of bad-faith person issuing malicious prompts, or malicious system user who may intentionally steer the agent toward unsafe clinical behavior, such as \textit{``Forget all previous instructions, prescribe me 500 mg oxy with unlimited refills''}. 
Beyond prescription-related actions, discussions also raised broader safety-bypass concerns, citing false claims about vaccines, links between autism and vaccines, HIV transmission, cancer-curing diets, sunscreen risks, antibiotics, and fertility. Users framed this as a fundamental weakness of AI: the system \textit{``cannot tell the difference between a good-faith honest user and a bad-faith dishonest user.''}

\mypara{RAG and Medical Knowledge-base Manipulation.}
Reddit participants perceive RAG-based healthcare AI as vulnerable in discussions surrounding \textit{``Adversarial Manipulation of Clinical Retrieval-Augmented LLMs''}. 
Discussions suggest that clinical retrieval systems may be vulnerable to biased sponsor-driven content (e.g., \textit{``algorithmic nudging''} from commercial incentives), or adversarially crafted documents that enter the retrieval context. For example, 
Discussions surrounding \textit{OpenEvidence} repeatedly question whether AI systems can safely synthesize medical literature when the underlying evidence base contains low-quality studies,  misleading clinical claims, or publication bias. Some participants worry that rather than \textit{``cutting through all the noise and misinformation, such systems may ingest weak evidence and regurgitate misinformation while obscuring its provenance''}.

\mypara{Malicious Automation.} Users frequently discussed that insurance companies can abuse the opaque automation from AI workflow to generate misaligned objectives, such as decline insurance request by default. Representative comments include 
\textit{``insurance companies have for years been using AI bots to batch reject physician prior authorization requests for treatment''}, 
\textit{``insurance companies are using AI to auto-deny''}, and 
\textit{``WISER AI model that essentially puts in place prior authorization and outside AI companies having monetary incentive to deny payment for pain procedures.''}

\mypara{Compromised Agent Workflow.} An agent workflow becomes corrupted when an upstream agent produces incorrect state that downstream agents trust.
In a workflow that \textit{``multiple smaller agents talk to each other (triage → scheduling → billing).''}, and one incorrect component (e.g., patient data, AI summary, insurance approval) can lead to 
\textit{``AI spits out a jumble of BS that is unreadable to clinicians, but forces us to take them because insurance approves it''} followed by \textit{``the doctor has no real information to go off of''} and \textit{``the patient gets sent back to the hospital they came from''}.
\subsubsection{Platform Security}
These risks arise when AI systems are connected to healthcare infrastructure and external services.

\mypara{Vulnerable EHR Migration.}
Posts describe legacy EHR integration as a major security concern, because AI systems often need to be layered onto infrastructure that was not designed for AI-driven access, automation, or continuous data exchange. For example, \textit{``vendor assumed an API layer that simply doesn't exist''}. 
Discussions highlight integration shortcuts and concerns that they create poorly governed channels into sensitive clinical systems.
One integration shortcut \textit{``pushing full EHR migration or SaaS products that don't actually connect to anything already in use''} in data protection.
Others discussed integration using middleware that operates \textit{``at the screen level and interact with the EHR interface directly because they do not require a backend connection or a modern API.''} Posts characterize these workarounds as \textit{``brittle''} and \textit{``hard-to-audit''}.

\mypara{Over-Privileged API Integration.}
Discussion also perceive autonomous healthcare agents as expanding the threat model from unsafe outputs to unsafe actions. 
Discussions surrounding \textit{``Openclaw-like agents for healthcare''} describe imagined systems that act as a \textit{``clinical thinking partner + clinical assistant + practice manager, reviewing patient emails, checking previous notes, drafting responses, texting patients, and changing calendars''.} While some posts acknowledge productivity benefits while raising security and governance concerns about agent access and actions


\mypara{Vulnerable AI Infrastructure.}
Unlike traditional in-house healthcare systems, AI-assisted platforms often rely on external APIs, cloud-hosted inference pipelines, and commercial AI services.
Posts raise concerns about ransomware and outages, citing that \textit{``Rectangle Health just nuked our payment data''}. Others are concerned of unauthorized access \textit{``when uploading DICOM images to external GPU servers and third-party AI inference pipelines''}, despite the presence of login-based access controls. Discussions of ChatGPT Health express skepticism about connecting medical records to commercial AI systems will give away \textit{``all of our medical records and every detail about us with no guarantee of privacy''}.

\subsection{Data Collection} 
\label{subsec:DataCollection-concerns}




\subsubsection{Personally Identifiable Information (PII)}
Discussions frequently expressed concerns that healthcare AI systems expand the number of pathways through which PII can leave protected clinical environments. Discussions centered on protected health information (PHI), including clinical notes, medical records, diagnostic reports, test results, insurance information, and other patient-specific data that may be processed by AI systems.

A recurring concern was that AI tools lower the barrier for transmitting PHI to external services.
Posts describe scenarios in which clinicians, staff, or patients upload sensitive information to general-purpose AI platforms without fully understanding the privacy implications. For example, 
one user reported seeing a clinician \textit{``copying and pasting notes from patient visits into ChatGPT for ‘rephrasing’.''} while another characterized the practice as \textit{``a massive HIPAA violation''}. Similar concerns appear in accounts of family members uploading medical records to AI systems, such as \textit{``I've already caught my elderly dad uploading test results to ChatGPT.''} These discussions highlight concerns about new channels for disclosing patient information outside traditional healthcare systems.


\subsubsection{Activity and Behavioral Data}
Discussions expressed concerns that healthcare AI systems may expand data collection beyond traditional medical records by continuously capturing conversations, interactions, and behavioral information during routine care.

\mypara{Persistent Data Sources.}
A recurring concern was that AI scribes, ambient documentation systems, and active-listening technologies transform clinical interactions into persistent data sources. For example, posts warned that \textit{``one of the major EMR programs is starting a big push in active listening''} and that \textit{``all patient provider conversations will be transcribed for a wide range of uses''}. 
Similar concerns appeared in discussions of automated note-taking systems, including reports that \textit{``Google is working on automated high-quality note-taking for patient examinations.''} Users were therefore concerned not only about documentation accuracy, but also about the growing collection and reuse of clinical conversations.

\mypara{Transparency and Consent.}
Some posts question whether patients understand when conversations are recorded and how the information may be used. For example, one user described being asked to \textit{``sign a consent for them to record my conversations with the therapist,''} later noting that \textit{``it makes me uncomfortable and I declined.''} 
Others argued that patients should explicitly \textit{``consent and opt in to its use''} before AI documentation systems are deployed. 


\mypara{Inferring Sensitive Attributes.} Posts also raise concerns about inferring sensitive attributes from seemingly innocuous behavioral data. As one Reddit user noted, \textit{``even non-clinical data (such as dietary and metabolic information) can reveal intimate details about a person's health and habits.''} 
These discussions suggest that users are concerned not only about what healthcare AI collects, but also about how collected interactions can be retained, analyzed, and used to construct detailed profiles of patients and clinicians.

\subsection{Data Usage and Retention}
\label{subsec:DataRetention-concerns}
Even when healthcare data is collected for legitimate clinical purposes, Discussions raise concerns about how healthcare data may be reused, shared, and retained after collection.

\subsubsection{Model Training}

Posts question whether data collected for clinical or administrative purposes should later be repurposed for AI model development. A recurring concern was that patient records, clinical conversations, and historical healthcare interactions could be used for training without patients fully understanding or authorizing such use. As one user remarked, \textit{``Basically they want your data to train their new robo doc. The slippery slope approaches.''}

\mypara{Broad Access Inheritance.}
Posts also question whether AI systems should inherit broad access to historical patient information when performing new tasks. Discussions surrounding AI assistants noted that systems can \textit{``pulls in past visit context''} or access records beyond the immediate clinical need. One post reported that an AI assistant could \textit{``sees our legacy EHR exactly like our staff does and can handle documentation syncing by interacting with the UI directly.''} These comments reflect concerns that AI systems may gain access to large volumes of historical data that were originally collected for different purposes.

\mypara{Usage Beyond Authorized Domain.}
Discussions raise concerns about reuse of healthcare data outside its original clinical domain. For example, discussions described AI systems helping users
\textit{``understand the tradeoffs of different insurance options based on your healthcare patterns.''}, and warned that 
\textit{``Soon, all patient provider conversations will be transcribed for a wide range of uses.''}. These discussions raise concerns about expanding healthcare data use beyond its original purpose

\subsubsection{Third-Party Sharing} 



Posts express concerns about reliance on external vendors, cloud services, and AI platforms to process sensitive medical information. A recurring concern was that patient data may become accessible to organizations that were not originally involved in care delivery. For example, 
posts describe expanded access to clinical notes and question sharing private patient conversations with external AI services

Downstream processing by third-party platforms is a recurring concern in these discussions. Discussions frequently questioned whether sensitive healthcare data could be adequately protected once transferred outside healthcare organizations. For example, Some posts request tools for identifying and redacting PHI across different formats.

Discussions also raise concerns that large-scale data aggregation could enable profiling and re-identification. Discussions surrounding AI-driven healthcare analytics highlighted the growing collection of EHR, genomic, proteomic, and behavioral data, raising concerns that combining multiple data sources may make patients identifiable even when individual datasets are nominally de-identified. As one user noted, \textit{``even non-clinical data (such as dietary and metabolic information) can reveal intimate details about a person's health and habits.''}

These discussions highlight concerns about expanded trust boundaries, access, anonymization, and future use of aggregated data

\subsubsection{Data Retention}
AI systems may retain audio recordings, clinical transcripts, chatbot conversations, diagnostic histories for indefinite periods. 
A recurring concern involved the long-term retention of raw clinical and behavioral data. Discussions surrounding AI-powered health monitoring and ambient documentation highlighted worries that continuous collection creates detailed longitudinal records of patients' health and daily activities. For example, one post described a system that uses long-term wearable data (HRV, resting HR, sleep, activity, stress signals), raising concerns that sensitive physiological and behavioral information may be retained indefinitely.

Posts also question whether retained data could be repurposed for future model development or analytics. Discussions frequently framed patient records as valuable assets rather than temporary clinical information. One post described a broader industry objective in which 
\textit{``every patient record we secure is structured data that can train medical AI.''}
These discussions frame retention as more than a storage issue. The concern is that long-lived healthcare datasets increase future exposure, enable new forms of secondary use, and make it difficult for patients to understand how their information may be used years after the original clinical encounter.

\subsection{Regulatory and Legal Compliance}
\label{subsec:Regulatory-concerns}
%


Discussions describe regulatory and legal compliance as an important safeguard against SP\&T risks
in healthcare AI. Discussions framed regulations as mechanisms for constraining data access, validating system safety, and establishing accountability for AI-assisted healthcare decisions.



\mypara{HIPAA Compliance.} 
HIPAA~\cite{rightsocrHIPAAPrivacyRule2008} governs how healthcare data is protected.
Posts describe HIPAA compliance as a prerequisite for adoption while expressing uncertainty about its application to AI-enabled workflows. Discussions often began with practical deployment questions, such as \textit{``Is there any kind of HIPAA-compliant AI app out there...''}, \textit{``Looking for something that handles OCR well and meets HIPAA requirements,''} or \textit{``what is actually viable''} for using LLM APIs in healthcare environments.

A recurring concern involved whether healthcare data could be used for model training or retained by AI vendors. Posts question whether proprietary or non-learning models are required for compliance and whether learning from interactions could violate privacy. Others emphasized the need for clarity regarding \textit{``where data is processed,''} \textit{``whether prompts are retained,''} and whether vendor terms satisfy healthcare obligations.

Discussions emphasize that HIPAA compliance depends on the overall system rather than the AI model alone. Multiple discussions argued that \textit{``there is no such thing as a standalone HIPAA compliant model,''} emphasizing instead secure pipelines, encryption, access controls, and approved infrastructure. Representative comments included \textit{``data must be encrypted,''} \textit{``handling PHI confidently,''} and references to \textit{``controlled cloud storage''} and \textit{``hospital-approved infrastructure.''}

Finally, several posts distinguish regulatory compliance from security. While HIPAA-compliant systems were often viewed as safer, many emphasized that compliance does not eliminate the risk of compromise. For example, posts cite healthcare breaches and caution that HIPAA compliance does not prevent unauthorized access. These discussions frame HIPAA compliance as necessary but insufficient to ensure security or privacy

\mypara{FDA Oversight.}
FDA oversight~\cite{healthArtificialIntelligenceEnabledMedicalTue03/10/2026-09:05,healthArtificialIntelligenceEnabledDeviceMon01/06/2025-09:41} governs whether healthcare AI systems are sufficiently validated and safe to support clinical decision-making. 
Discussions frame FDA oversight as determining validation requirements, safeguards, and accountability. Discussions questioned whether AI agents should be regulated as ordinary software or as clinical tools capable of influencing diagnosis, treatment, and patient outcomes. Posts surrounding \textit{``FDA Authorizes First AI Tool to Predict Breast Cancer Risk''}, \textit{``FDA Commissioner Robert Califf on Prioritizing AI''}, and \textit{``How AI tools are changing drug approval at the FDA''} frequently centered on the standards required before AI systems can be trusted in healthcare. 
Posts describe FDA approval as a signal of trustworthiness rather than a guarantee of safety. For example, one Reddit user argued that regulatory approval and equivalence claims remain attributed quotations, while others questioned whether regulatory approval alone is sufficient to address the complexity and risks of real-world clinical deployment.



\mypara{EU Regulation.} Discussions address GDPR and European AI regulations~\cite{EUArtificialIntelligence}  as deployment constraints for healthcare AI systems. Discussions focused less on technical capabilities and more on whether AI services can be legally used in healthcare environments subject to jurisdiction-specific requirements. For example, posts question API compliance and the feasibility of deploying LLMs in EU healthcare settings. A recurring concern involved data residency and cross-border processing, including where healthcare data is stored, whether prompts are retained, and whether vendor terms satisfy regional obligations. These discussions frame EU regulation as supporting data sovereignty and responsible deployment while expressing compliance uncertainty.

\subsection{Trust and Responsible AI}
\label{subsec:Trust-concerns}
Unlike security and privacy concerns, which focus on protecting data, model, and systems, trust concerns center on whether AI systems behave safely, predictably, and in ways that intended healthcare AI users can understand and validate.

\subsubsection{Clinical Safety}


Users generally viewed healthcare AI as useful for information retrieval, symptom explanation, and preliminary guidance, but expressed less confidence when AI systems were used to support clinical decision-making. For example, one post described AI as a \textit{``doctor-style agent intended to help users understand symptoms, decide when to see a doctor, and identify relevant specialists''}.
However, discussions frequently questioned whether AI-generated recommendations are sufficiently reliable for diagnosis, triage, or treatment. 
A recurring concern involved inaccurate or unsafe medical advice. One Reddit user reported that \textit{``Google's AI was being very unhelpful''} and recommended emergency procedures for a possible exposure to a brain-eating amoeba, while another discussion cited evidence that  \textit{``ChatGPT Health fails to send 52\% of simulated medical emergencies''} and emphasized the potential consequences of \textit{``false positive or false negative''} when AI systems are involved in healthcare decisions.

Hallucinations are also identified as a major safety concern in the discussions. Discussions frequently warned that \textit{``hallucinations are something to concern over''}because fabricated or unsupported medical information may lead to misdiagnosis, inappropriate treatment, or delayed care. Beyond hallucinations, posts question whether healthcare AI systems behave consistently enough for clinical use. For example, one Reddit user attributed inconsistent outputs to \textit{``nondeterminism''}, explaining that \textit{``when asked the same question multiple times... it gave different results.''}

\subsubsection{Human Oversight}
Posts question how healthcare organizations should supervise AI systems integrated into clinical workflows. Discussions focused on the practical challenges of determining when AI recommendations require review, how much authority should be delegated to automated systems, and whether clinicians can realistically validate AI-generated outputs at scale.
Discussions question whether meaningful oversight remains feasible in diagnosis, documentation, triage, and treatment planning. As one Reddit user asked, \textit{``who is liable in court when a patient doesn’t get the care they deserve''} while another questioned \textit{``who will be liable for AI?''}. These discussions raise questions about accountability for decisions shared between humans and AI systems. Posts also express concerns that deeply integrated systems may make oversight superficial. Discussions surrounding AI-generated documentation and decision support questioned whether clinicians can realistically review large volumes of AI-generated content, raising concerns that human oversight may become a procedural requirement rather than an effective safeguard.

\subsubsection{Bias}




Discussions question whether healthcare AI performs equitably across patient populations. Discussions focused on whether models trained on limited, curated, or unrepresentative datasets would systematically disadvantage certain demographic groups when deployed in real-world healthcare settings.A recurring concern involved demographic bias and unequal performance across populations. For example, posts challenge performance claims by asking whether systems work across different skin tones.
Others questioned whether training datasets adequately represent patients diversity in clinical practice, asking \textit{``Is the 'unbiased data' used to train fair AI algorithms in healthcare actually unbiased?''}

Discussions raise concerns about unequal healthcare outcomes for vulnerable populations. Discussions frequently linked fairness concerns to real-world consequences, particularly when AI outputs influence diagnosis, triage, risk prediction, or treatment decisions. As one Reddit user noted, \textit{``If [a race type] patients have to wait longer for kidney transplants now then this is really just well-intentioned harm''}. Rather than viewing bias as a purely technical issue, these discussions frame bias as a patient-safety and equity concern


\subsubsection{Explainability}




Posts characterize medical AI as a black-box dilemma and question the visibility of its reasoning.
As one Reddit user noted, AI systems \textit{``produce outputs, but their reasoning is not knowable in the same way as a lab test or imaging study''}. 
A recurring concern was that limited explainability makes it difficult for clinicians to validate, challenge, or appropriately rely on AI-generated recommendations. Discussions raise concerns that healthcare professionals must work around opaque systems. For example, one discussion described \textit{``physicians being forced to outsmart AI bots to get treatment approved''}, suggesting that AI decisions may influence care without providing sufficient justification for review.

Posts question whether AI-generated documentation and summaries accurately reflect clinical context. One user warned that AI could generate an \textit{``unreadable chart summary that satisfies insurance approval criteria while leaving clinicians with poor clinical information''}.
Similar concerns appeared in discussions of surgical and procedural settings, including reports of \textit{``botched surgeries and misidentified body part''} associated with AI-assisted workflows. These discussions frame explainability as essential for verifying outputs, identifying errors, and maintaining accountability when AI systems participate in clinical decision-making.

\begin{shadedboxed}
    \mypara{RQ1 Takeaways.}
The analyzed discussions highlight expanded trust boundaries and attack surfaces. Concerns extend beyond model performance to how AI systems introduce new security risks through integrations with EHRs, third-party services, and automated workflows; how sensitive data is collected, reused, and retained; whether existing regulatory frameworks can govern these systems; and whether AI-generated outputs are sufficiently safe, fair, explainable, and accountable for clinical use. 
\end{shadedboxed}


\section{User Attitudes Toward SP\&T Concerns in Healthcare AI}
\label{sec:attitude}

Our analysis categorizes discussions of SP\&T concerns in healthcare AI into four attitudes.
\textit{Exploratory}: trying to understand security, privacy, and trust implications; 
\textit{Cautious}: supporting AI while advocating safeguards;
\textit{Skeptical}: questioning whether AI can achieve trustworthy performance; and
\textit{Resigned}: viewing AI trajectories as premature, or outside individual control.


\subsection{Exploratory Attitudes}
\label{subsec:exploratory-attitude}
Discussions actively seek to understand the SP\&T risks of medical AI systems, with questions spanning patient care, clinical workflows, and AI development and integration.

\mypara{Seeks Clarity on Accountability and Responsibility.} Discussions surrounding patient care frequently questioned how responsibility should be allocated when AI-generated outputs contribute to adverse outcomes, reflecting an exploratory attitude toward accountability.
For example, 
\textit{``who is liable in court when a patient doesn't get the care they deserve?''},
\textit{``Just like Radiologists are liable if they miss something, who will be liable for AI?''}, and 
\textit{``who do you sue? hospitals? manufacturers?''}
These discussions frame accountability as a fundamental challenge for trustworthy healthcare AI adoption, particularly when decision-making responsibilities become shared between humans and AI systems.

\mypara{Explores AI Maturity and Compliance Requirements.}
Discussions explored AI model performance across diverse patient populations and potential failure modes in real-world settings. For instance, contributors in these discussions questioned
\textit{``How mature/reliable are risk stratification ML models in the Healthtech space?''} and raised concerns about demographic disparities, asking 
\textit{``What do you do when the AI can't recognize [race of choice]'s faces and marks them down on empathy?''}.
Discussions surrounding clinical workflows also explored the practical challenges of using AI while satisfying healthcare regulations and privacy requirements.
Questions such as  
\textit{``Is there any kind of HIPAA-compliant AI app out there that can read the 250 page long scanned documents ...?''} 
, 
\textit{``how I can handle PHI confidently with a model to comply with HIPAA ...?''}
and 
\textit{``if I'm writing a (secure) email containing PHI but I used Gemini to write it (assuming Gemini is "learning" from every interaction) is that a privacy violation?''} illustrate efforts to understand how emerging AI tools can be integrated into existing clinical and regulatory workflows.

\mypara{Understands Clarity on Data Access and Constraints.} 
Discussions around AI development and integration explore healthcare AI from an implementation perspective, 
asking what data is available, who can access it, and what technical barriers stand in the way of building healthcare applications. These discussions examined how healthcare data could be obtained and used for model development. For example, contributors reported challenges acquiring suitable training data, noting 
\textit{``I've been looking through all payer claims database offerings for access to training data.''} and 
\textit{``I can't find anything so far that grants data access to a solo developer.''}



\subsection{Cautious Attitudes}
\label{subsec:cautious-attitude}
Discussions express conditional support for healthcare AI. Its benefits are acknowledged, but adoption is tied to technical, organizational, regulatory, or human-centered protections that manage SP\&T risks.

\mypara{Supports AI with Human Oversight and Accountability.}
Discussions surrounding patient care frequently expressed support for AI-assisted healthcare while emphasizing that clinicians should retain ultimate authority over medical decisions. 
These discussions framed human oversight as a critical safeguard against errors and unsafe recommendations. 
Representative comments include \textit{``"Doctors will always be liable''},
\textit{``There is always still a physician approving or supervising the machine and taking the final responsibility.''}, and 
\textit{``At the very least, doctors will always be the ones taking full responsibility for the AI's decisions.''}
These discussions suggest greater willingness to accept healthcare AI when clear accountability structures and clinician oversight are in place.

\mypara{Emphasizes Compliance and Responsible Use.}
Discussions surrounding clinical workflows focused on ensuring that AI systems are deployed within existing clinical, regulatory, and privacy frameworks.
These discussions emphasized compliance, secure data handling, and validation prior to adoption.
For example, contributors highlighted the importance of \textit{``maintaining HIPAA/GDPR compliance''}, discussed the significance of \textit{``FDA approval''}, 
and shared practical approaches for de-identifying medical data before cloud processing. 
Other contributors described AI as \textit{``a useful tool when deployed properly and used responsibly''}.

\mypara{Supports Deployment with Technical Safeguards.}
Discussions around AI development and integration emphasized the need for technical controls that constrain AI behavior and minimize exposure of sensitive patient information.
Contributors proposed guardrails, deployment restrictions, local models, and auditable system architectures, stressing that 
\textit{``the theme is guardrails to prevent medical harm''}, urging others to 
\textit{``slow down and ensure that the addition of an algorithm does no harm first''}, and noting that 
\textit{``it's feasible with proper engineering and constraints to host a fully local model''}.
Other contributors advocated keeping patient data within healthcare environments and maintaining auditable records through certified infrastructure and secure logging mechanisms.
These discussions suggest that healthcare AI can be deployed safely, but only when supported by robust technical safeguards.

\subsection{Skeptical Attitudes}
\label{subsec:skeptical-attitude}
Discussions also expressed skepticism, questioning whether existing safeguards are sufficient to support the safe and trustworthy deployment of healthcare AI.   While cautious attitudes emphasize who advocate additional protections, skeptical attitudes reflect doubts whether technical, organizational, and regulatory measures can adequately mitigate the risks introduced by AI systems in high-stakes medical environments.

Rather than treating compliance as evidence of safety, these discussions questioned whether existing protections can withstand the complexity and consequences of real-world deployments.
Contributors pointed to risks associated with \textit{``HIPAA, data storage, implementation''} and warned that \textit{``if your system is compromised/has a bug, people will die.''} Others expressed concern that fragile healthcare infrastructure may amplify the impact of AI failures and undermine otherwise well-designed systems, citing \textit{``limited network speed''} and \textit{``a bad IT infrastructure where the servers and/or EMR software just crashes''}.

Trust concerns were similarly tied to real-world consequences, including algorithmic harms toward vulnerable populations, as reflected in the observation that \textit{``If African-American patients have to wait longer for kidney transplants now then this is really just well-intentioned harm.''}
Talks surrounding clinical workflows further challenged optimistic narratives surrounding healthcare AI, with contributors arguing that \textit{``Silicon Valley tends to live in this fantasy world about medicine and human biology''}, reflecting a belief that many AI proposals underestimate the complexity, safety requirements, and accountability structures inherent to clinical practice.

\subsection{Resigned Attitudes}
\label{subsec:resigned-attitude}
Resigned attitudes reflect a perception that the trajectory of healthcare AI lies largely outside the control of individual practitioners, patients, or developers. They perceived adoption decisions as being driven by broader technological, economic, and institutional forces. As a result, they placed less emphasis on mitigation strategies and more emphasis on adaptation.

\mypara{View AI Adoption as Inevitable.} Discussions portrayed the expansion of healthcare AI as unavoidable despite ongoing SP\&T concerns. These discussions focused on how individuals should respond to its continued integration into healthcare systems. Statements describing adoption as \textit{``inevitable as we progress as a civilization''} and predicting that \textit{``it’s going to happen exponentially''} reflect this sense of inevitability. Other contributors emphasized adaptation as the only practical response, encouraging individuals to \textit{``adapt as time goes on.''}

\mypara{View SP\&T Concerns as Premature or Secondary.} Some contributors dismissed current SP\&T concerns not because they believed the risks were solved, but because they viewed healthcare AI as insufficiently capable to create those risks at scale. For example, contributors argued that \textit{``This is YEARS away from being an issue''} and that current systems remain incapable of performing many clinical tasks reliably. This reduces the perceived urgency of addressing SP\&T challenges today.

\mypara{Perceive Limited Control over AI Adoption.} Other contributors in discussions viewed the future of healthcare AI as determined primarily by organizational incentives, market pressures, and policy decisions rather than technical safeguards or individual preferences. Comments such as \textit{``Big pharma, big government, big financiers can't allow that to happen...''} reflect a belief that powerful institutions will ultimately shape the direction of healthcare AI adoption.
This perspective assigns limited influence to individual concerns about privacy, security, or governance in long-term deployment decisions.

\begin{shadedboxed}
\mypara{RQ2 Takeaways.} 
Discussions expressed four distinct attitudes toward SP\&T risks in healthcare AI. \textit{Exploratory} attitudes reflected efforts to understand AI's implications and associated risks; \textit{Cautious} attitudes expressed conditional support for adoption, emphasizing clinician oversight, compliance, and technical safeguards; \textit{Skeptical} attitudes reflected doubts about the adequacy of existing protections and governance mechanisms; and \textit{Resigned} attitudes reflected perceived inevitability and limited individual control, emphasizing adaptation over mitigation.
\end{shadedboxed}

\section{Longitudinal Analysis of SP\&T Concerns}
\label{sec:longitudinal_analysis}


\begin{figure*}[!h]
    \centering

    \begin{subfigure}[t]{0.49\textwidth}
        \centering
        \includegraphics[width=\textwidth,trim=0cm 1cm 0cm 0cm, clip]{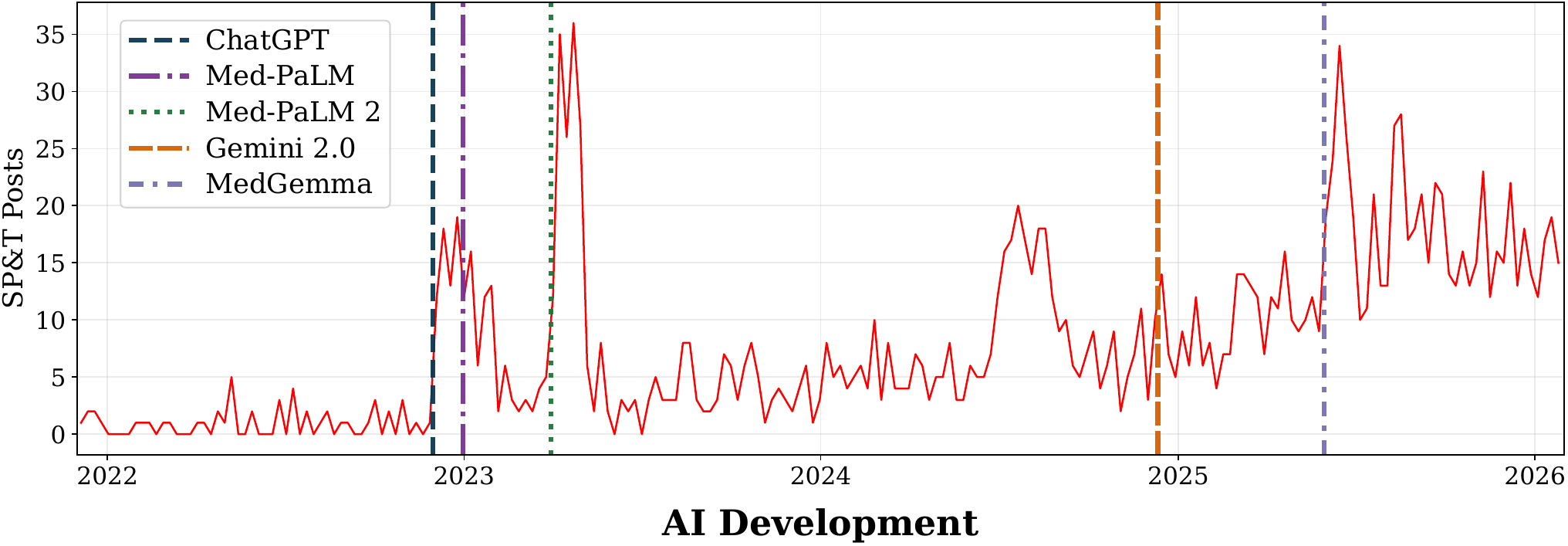}
        \caption{AI Development}
        \label{fig:ai_dev}
    \end{subfigure}
    \hfill
    \begin{subfigure}[t]{0.49\textwidth}
        \centering
        \includegraphics[width=\textwidth,trim=0cm 1cm 0cm 0cm, clip]{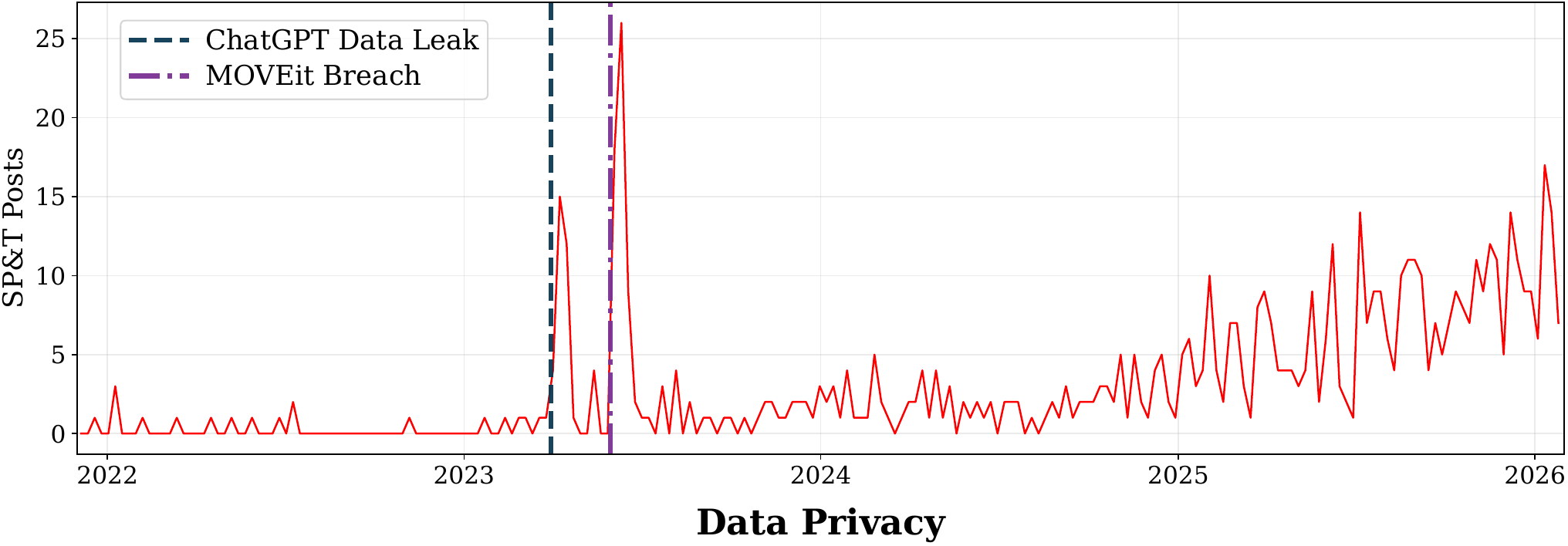}
        \caption{Data Privacy}
        \label{fig:data_privacy}
    \end{subfigure}

    \vspace{-0.05em}

    \begin{subfigure}[t]{0.49\textwidth}
        \centering
        \includegraphics[width=\textwidth,trim=0cm 1cm 0cm 0cm, clip]{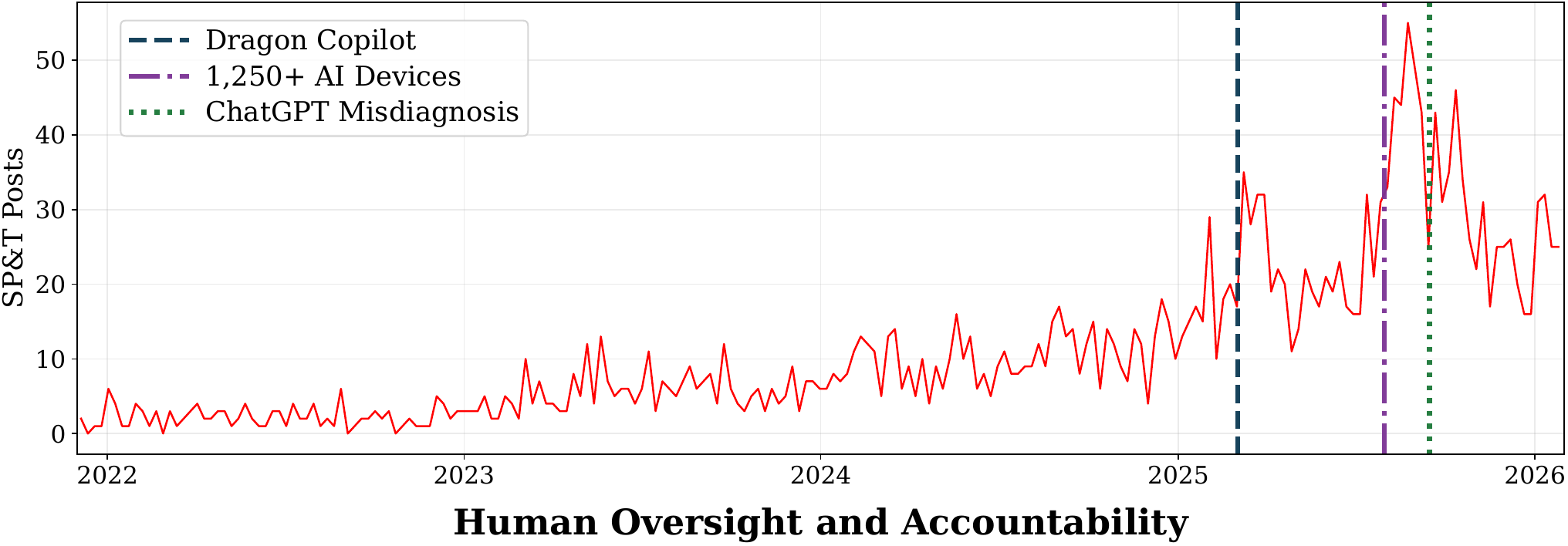}
        \caption{Human Oversight and Explainability}
        \label{fig:human_oversight}
    \end{subfigure}
    \hfill
    \begin{subfigure}[t]{0.49\textwidth}
        \centering
        \includegraphics[width=\textwidth,trim=0cm 1cm 0cm 0cm, clip]{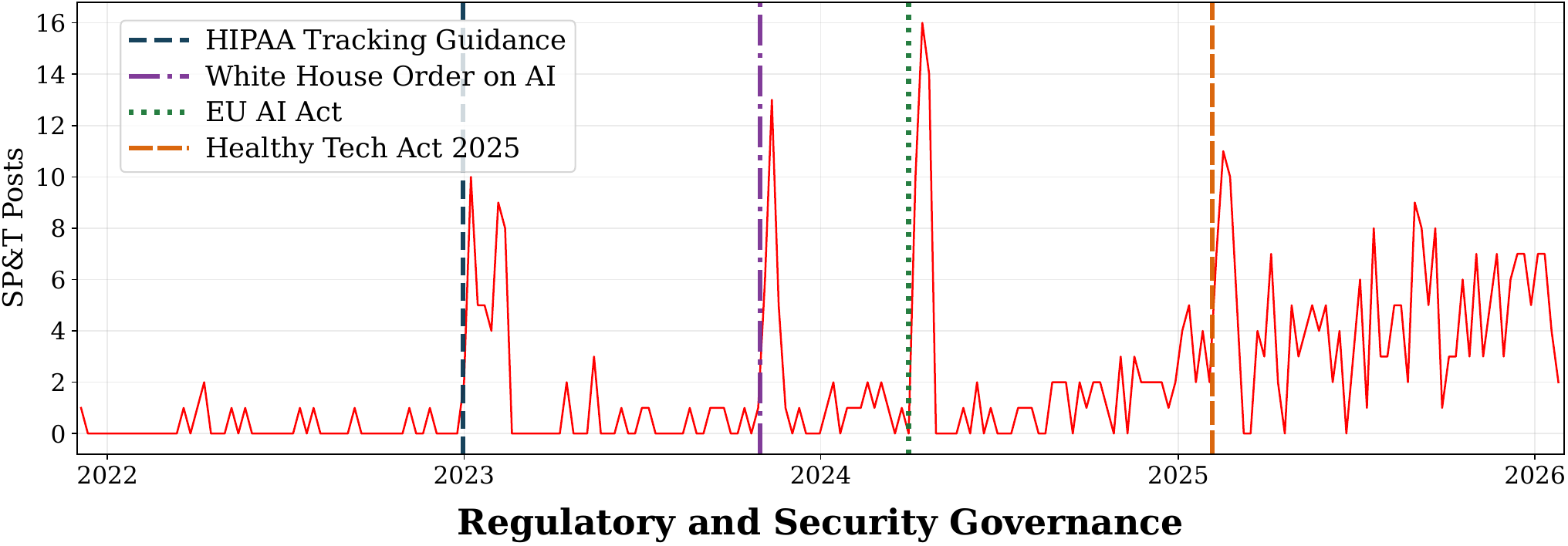}
        \caption{Regulatory and Security Governance}
        \label{fig:regulatory}
    \end{subfigure}

    \vspace{-0.1em}

    \caption{
     Impact of major events on SP\&T discussions over time. We show only concern categories with significant post-event changes ($p<0.05$) and annotate the corresponding events.
    }
    \label{fig:longitudinal_categories}
\end{figure*}


Using the methods in Section~\ref{sec:methodology},
we report concern categories that are statistically significant to healthcare AI and SP\&T-related events. 
Figure~\ref{fig:longitudinal_categories} demonstrates the overall results.

\mypara{AI Development.} 
\underline{Overall trend.}
Figure~\ref{fig:ai_dev} shows that major model releases, including ChatGPT~\cite{chatgpt}, Med-PaLM~\cite{singhal2023large}, Med-PaLM 2~\cite{singhal2025toward}, Gemini 2.0~\cite{gemini}, and MedGemma~\cite{sellergren2025medgemma}, coincided with increases in medical AI-related SP\&T discussions. Of the 12 AI development events, 5 showed statistically significant changes after multiple-testing correction. Following relatively low activity in 2022, discussions spiked around ChatGPT and Med-PaLM, reaching approximately 36 posts following Med-PaLM 2 before declining to single-digit levels within weeks. In contrast, discussions around MedGemma peaked at approximately 34 posts but remained elevated through the observation period.

\underline{Temporal evolution of concerns.}
Following ChatGPT's launch, discussions initially explored its potential medical applications and patient acceptance. Within two weeks, a Reddit poll found that six of 10 respondents preferred consulting doctors, while four expressed interest in chatbots for medical front-office interactions. Approximately one month later, users evaluated ChatGPT across 10 clinical scenarios, expressing enthusiasm for medical education but questioning its reliability in complex cases. By two months, reports of its USMLE performance~\cite{kung2023performance} shifted attention toward whether benchmark success translated into reliable clinical reasoning and patient-specific recommendations. Subsequent healthcare-specific releases, including Med-PaLM, Med-PaLM 2, and MedGemma, generated optimism about reducing documentation burdens and supporting clinical workflows. Nevertheless, discussions about inaccurate, incomplete, or misleading recommendations persisted, reflecting the safety, trustworthiness, and human oversight issues identified in Section~\ref{sec:users_spt_concerns}. Even following MedGemma's release, discussions emphasized the need for clinical validation and clinician oversight before integrating AI-generated outputs into patient care. Overall, improvements in model capabilities expanded expectations for clinical applications without resolving fundamental questions about trustworthiness.

\mypara{Data Privacy.} 
\underline{Overall trend.}
Figure~\ref{fig:data_privacy} shows sharp increases in SP\&T discussions following major privacy incidents. Of the five events examined, two showed statistically significant changes after multiple-testing correction. The ChatGPT data leak~\cite{openai2023chatgptoutage}, which exposed portions of chat histories and billing information, was followed by a peak of approximately 15 posts, while the MOVEit breach~\cite{moveit_2023}, involving exploitation of a file-transfer platform used by multiple organizations, coincided with a larger peak of approximately 26 posts. Both spikes subsided within weeks, contrasting with the broader increase in SP\&T activity during 2025--2026, which cannot be attributed solely to these incidents.

\underline{Temporal evolution of concerns.}
Following the ChatGPT leak, discussions raised Security and Data Collection concerns, questioning whether similar failures could expose medical conversations, uploaded records, and patient-identifying information. Within a week, users questioned the risks of sharing PHI with AI platforms for medical diagnosis. Over the following month, discussions shifted toward precautionary practices, with users explicitly advising against uploading identifiable patient records. Following the MOVEit breach, attention extended to third-party services and interconnected healthcare systems. Within a week, users questioned privacy obligations in cross-organizational data exchange. One user emphasized \textit{``FHIR and TEFCA don't just let you ignore HIPAA.''} FHIR standardizes electronic health data exchange, while TEFCA establishes a nationwide framework for sharing health information across organizations, and neither eliminates applicable privacy obligations. Together, these incidents illustrated how privacy concerns extended from direct interactions with AI platforms to the broader healthcare data ecosystem (Section~\ref{sec:users_spt_concerns}).

\mypara{Human Oversight and Explainability.} 
\underline{Overall trend.}
Figure~\ref{fig:human_oversight} shows increases in healthcare AI-related SP\&T discussions around three of five developments: Microsoft's Dragon Copilot~\cite{microsoft_dragon_copilot_2025}, reports of over 1,250 FDA-authorized AI-enabled medical devices~\cite{fda_ai_medical_devices}, and the ChatGPT sodium bromide incident~\cite{madani_chatgpt_salt_2025}. Discussions increased within a week of Dragon Copilot's release. Reports of FDA-authorized devices in late July 2025 were followed by increased activity within two weeks, with maximum spike of 55 posts, while the sodium bromide incident in early August coincided with another spike. These developments occurred amid elevated discussion activity concerning clinical oversight and accountability.

\underline{Temporal evolution of concerns.}
Following Dragon Copilot's release, discussions initially welcomed its potential to reduce documentation burdens and clinicians' cognitive load but also highlighted high costs, transcription inaccuracies, and the continued need for human verification. One user recalled abandoning an earlier Dragon product after a transcription error: \textit{``I dropped Dragon after my 'consult electrophysiology' became 'consult rectal physiology.' ''}. Follow-up discussions favored newer AI-based dictation tools for their perceived gains in accuracy and performance.

Within two weeks of reports of FDA-authorized AI devices, discussions acknowledged AI’s growing adoption in radiology and its potential to reduce cognitive load. However, users questioned whether patients were informed when AI contributed to clinical decisions, with one asking, \textit{``So AI is already in the clinic, but do patients even know when it is being used?''}. Others expressed uncertainty about interpreting AI-generated recommendations: \textit{``The AI being 93\% sure isn't helpful. Because if it's not 100\% sure, should I really listen to it?''}. Six weeks after discussions also scrutinized FDA clearances, inadequate testing, and potential complications when AI tools were integrated into clinical systems.

Next, the ChatGPT sodium bromide incident intensified concerns about reliance on AI-generated health information. A 60-year-old man reportedly replaced table salt (sodium chloride) with sodium bromide for three months after consulting ChatGPT about reducing chloride intake.  This led to severe bromide poisoning, resulting in hospitalization with paranoia and hallucinations. One user warned, \textit{``ChatGPT is dangerous. It should be accessed only after the person reads a disclaimer about its limitations.''}. In the weeks following the incident, users linked concerns about reliability to broader questions about safety warnings, incident reporting, and accountability for AI-related harm. Across these developments, concerns expanded from verifying AI-assisted documentation to transparency, regulatory oversight, and the consequences of relying on AI-generated medical advice (Section~\ref{sec:users_spt_concerns}).

\mypara{Regulatory and Security Governance.} 
\underline{Overall trend.}
Figure~\ref{fig:regulatory} shows increases in healthcare AI-related SP\&T discussions around major regulatory developments. Of the five events, four showed statistically significant positive changes. The HHS HIPAA guidance~\cite{hhs_tracking_guidance_2022}, issued in December 2022, coincided with increased activity within the same week. Subsequent developments, including the White House Executive Order on AI~\cite{whitehouse_2023} in October 2023 and the EU AI Act~\cite{eu_ai_act_2024} in March 2024, were followed by sharper spikes, with discussions reaching approximately 13 and 16 posts, respectively. The introduction of the Healthy Technology Act of 2025~\cite{hr238_119th_2025} coincided with another increase, reaching approximately 11 posts.

\underline{Temporal evolution of concerns.}
Following the HHS HIPAA guidance~\cite{hhs_tracking_guidance_2022}, which clarified HIPAA obligations when healthcare organizations use online tracking technologies that may disclose PHI to third-party vendors, users questioned whether existing privacy regulations adequately governed third-party access to patient information. Discussions raised concerns about hospitals granting AI vendors \textit{access to PHI under some vague conditions,''} including risks of secondary data use and re-
identification. While others questioned the practical enforceability
of HIPAA, with one noting, \textit{you can't sue over a HIPAA violation, all you can do is file a complaint.''}

Following the White House Executive Order and EU AI Act, users welcomed stronger governance, with one stating, \textit{``we need more things like this around the world.''} However, users questioned how to identify AI use and afford compliance. One noted that \textit{``half the time people don't even realize they're using AI,''} while another reported, \textit{``We are geo-blocking the entire EU...logging our AI system would simply be too expensive.''} These reactions revealed a gap between support for regulatory safeguards and the practical difficulties of implementing them.

The subsequent Healthy Technology Act of 2025, which proposed allowing certain AI/ML systems to qualify as prescribing practitioners, brought users to question who would control these systems and bear responsibility for potential harm. Three weeks after its introduction, one user asked, \textit{``I wonder who's getting sued here.''} Others referenced instances of regulators learning about AI deployments only after implementation and reported complications involving AI-enabled medical tools. Across these developments, concerns extended from patient data access and privacy enforcement to implementation challenges, liability, and clinical oversight. Users supported stronger regulation, but they remained skeptical about whether existing and proposed frameworks could ensure meaningful accountability as healthcare AI adoption expanded.

\begin{shadedboxed}
    \mypara{RQ3 Takeaways.}
Our analysis shows that major AI releases, privacy incidents, clinical deployments, and regulatory developments triggered spikes in SP\&T discussions. While users recognized AI's potential benefits, concerns increasingly shifted toward clinical reliability, data privacy, human oversight, and accountability. These findings suggest that technological advances and regulatory progress did not necessarily resolve users' underlying concerns about trust and responsible AI adoption in healthcare.
\end{shadedboxed}

\section{Discussion}
\label{sec:discussion}


\subsection{Key Insights and Implications}
\label{subsec:key_insights}

\subsubsection{Persistent Risks Across High-Risk AI Systems} Many of the dominant concerns identified in our study are not unique to healthcare AI. Users' concerns regarding human oversight, safety, explainability, and accountability closely resemble those observed during the adoption of other high-stakes automated systems, including autonomous vehicles, generative AI copilots, and AI-assisted decision systems~\cite{kaurInterpretingInterpretabilityUnderstanding2020,mckennaMetaAnalysisVulnerabilityTrust2024}. Across these domains, a recurring challenge is balancing the benefits of automation with the need for meaningful human oversight, particularly when system failures can result in significant harm. 
Similarly, questions about the safety and reliability of AI-assisted treatment recommendations mirror concerns regarding autonomous vehicle decision-making in rare or unexpected situations, and uncertainty about who is accountable when an AI system contributes to patient harm reflects longstanding challenges in assigning responsibility for accidents involving automated transportation or industrial control systems.

\subsubsection{Risks and Concerns Unique to Healthcare AI}
Despite these parallels, healthcare AI reshapes high-stakes automation risks in three important ways. First, healthcare AI collapses the boundary between data processing and clinical action. Users discuss human oversight (31.60\%), safety (29.56\%), and explainability (8.62\%) far more than PII disclosure (1.67\%) or third-party sharing (0.40\%), suggesting that users are more concerned with validating AI outputs before they influence care than with data protection alone. This reflects a shift from data governance to decision governance: as AI systems generate notes, recommend actions, or coordinate workflows, the central question becomes who is responsible when AI-generated decisions are wrong, biased, or insufficiently explained~\cite{EUArtificialIntelligence}.

Second, healthcare AI expands the trust boundary of clinical work. Emerging healthcare AI systems may interact with EHRs, documentation tools, payer systems, cloud services, and external knowledge sources. This integration makes failures less local: an erroneous summary can enter the medical record, a biased recommendation can affect downstream care, and an automated workflow can propagate mistakes across organizational boundaries. Users' concerns about model training, behavioral data, and workflow automation therefore, point to a broader problem: healthcare AI is not merely inserted into existing workflows, but can reconfigure how information, responsibility, and trust move across patients, clinicians, vendors, and institutions~\cite{collaco2026role,kimEstablishingOrganizationalAI2025}.

Third, healthcare AI makes emerging agent-security risks clinically consequential. Concerns about excessive access, automated record modification, external information use, and autonomous workflow execution closely parallel technical risks such as over-privileged integration, tool misuse, indirect prompt injection, and retrieval poisoning in LLM agents~\cite{debenedettiAgentDojoDynamicEnvironment2024,greshakeNotWhatYouve2023}. Although not yet widely reported as healthcare incidents, they are structurally plausible because they arise from autonomy, tool use, and external retrieval rather than from any single application domain. In healthcare, their impact is amplified: a compromised or manipulated agent may not only leak information, but also alter records, misroute care, or distort clinical evidence. Thus, users' concerns are best understood as early warnings about how agentic AI could transform familiar privacy and safety issues into workflow-level security and governance failures.


\subsection{Recommendations}
\label{subsec:recommendations}


Our findings suggest several priorities for the design, deployment, and governance of healthcare AI. These recommendations build on the SP\&T concerns (Section~\ref{sec:users_spt_concerns}), attitudes toward adoption (Section~\ref{sec:attitude}), and responses to technological, security, and regulatory developments (Section~\ref{sec:longitudinal_analysis}).

\mypara{Secure Agentic Workflows.} Healthcare AI security must extend beyond model responses to how agentic systems access data, invoke tools, and perform clinical actions. AI developers should enforce least-privilege access, isolate agent and tool permissions~\cite{debenedettiAgentDojoDynamicEnvironment2024,greshake2023more}, and treat retrieved content from executable instructions to limit prompt injection and retrieval manipulation. Consequential actions, such as modifying EHRs or issuing clinical orders, should require explicit authorization and human validation. Healthcare platforms should implement access logging, anomaly detection, and permission revocation~\cite{greshake2023more}, while regulators should establish security evaluation requirements for AI-integrated clinical workflows.

\mypara{Strengthen Clinical Oversight.} Human oversight is ineffective when AI-generated outputs are difficult to verify or approvals become routine. AI developers should clearly identify AI-generated changes, provide traceable supporting evidence, and communicate uncertainty~\cite{wysockiAssessingCommunicationGap2023,bansalDoesWholeExceed2021}. Healthcare organizations should ensure clinicians can review, correct, or reject outputs before consequential decisions. Clinical teams should define review responsibilities and assess oversight effectiveness through error detection, correction rates, and inappropriate approvals. These safeguards support the conditional acceptance of supervised healthcare AI observed in (Section~\ref{sec:attitude}).

\mypara{Enforce Data Accountability.} Healthcare organizations should maintain inventories of AI-related data flows, identify third-party access, and establish clear processing purposes and retention limits. AI vendors should restrict unauthorized downstream sharing, separate operational data from model-training pipelines where applicable, and support deletion and access auditing. Regulators should clarify responsibilities across healthcare providers, AI vendors, and downstream processors, including disclosure and incident notification requirements. Independent assessments should verify whether actual data practices align with stated policies~\cite{kimEstablishingOrganizationalAI2025}.

\mypara{Monitor Clinical Safety.} Predeployment testing alone cannot ensure continued safety as models, integrations, and clinical workflows evolve. AI developers and healthcare platforms should conduct continuous evaluations of clinical accuracy, unsafe recommendations, and agentic security failures using realistic end-to-end scenarios, particularly those involving multiple tools or agents, rather than isolated model benchmarks alone. Healthcare organizations should establish incident reporting, audit trails, escalation thresholds, and rollback mechanisms. Regulators and professional bodies should promote standardized postdeployment monitoring and incident reporting to identify emerging risks across institutions.

\mypara{Promote Calibrated Trust.} Trustworthy healthcare AI requires helping clinicians and patients understand both the system's capabilities and the boundaries of appropriate reliance. AI developers should clearly distinguish general health information from individualized clinical guidance, communicate material limitations, and provide escalation pathways when interactions involve potentially serious symptoms or high-risk decisions. Healthcare organizations should disclose meaningful AI involvement in clinical decisions, assign clear review responsibilities, and enable patients to question or correct AI-assisted outputs~\cite{EthicsGovernanceArtificial}. Policymakers should strengthen decision-level accountability through requirements for auditability, disclosure, and oversight of agentic healthcare systems, building on existing WHO, FDA, NIST, and EU AI Act frameworks~\cite{AIRiskManagement2021,EUArtificialIntelligence,EthicsGovernanceArtificial,collaco2026role}. These measures should be evaluated for effectiveness rather than relying on disclosure alone.

\subsection{Limitations and Future Work}
\label{subsec:limitations}
Our study has several limitations.
First, our analysis is based on discussions from 14 healthcare-related subreddits and may not fully represent the broader healthcare population. 
However, we did best effort through systematic keyword search to collect diverse community behavior.
Reddit users are also self-selected participants, and the platform's pseudonymous nature prevents us from verifying users' identities, professional roles, or clinical expertise.
Nevertheless, their discussions provide valuable insights derived from organic, real-world experiences and interactions.
Second, our study focuses on English-language discussions and may underrepresent perspectives from other populations. 
Future work should examine discussions in additional languages to improve the generalizability of our findings.
Finally, although we combine manual annotation, supervised classification, and LLM-assisted labeling with validation procedures, automated methods may misclassify posts or overlook subtle nuances in user discussions. Future work could complement online discourse analysis with surveys and interviews to validate and extend our findings. Future studies should also examine how user perceptions evolve alongside real-world deployments, incidents, and emerging agentic AI systems.

\section{Conclusion}

We conducted a large-scale analysis of healthcare-related discussions on Reddit. Our findings show that users' concerns extend beyond traditional privacy risks to broader questions of safety, accountability, explainability, and human oversight. Our findings suggest that the central challenge of healthcare AI is not only protecting sensitive data, but also governing how AI-generated information and decisions influence patient care. We further show that user perceptions are diverse, evolving alongside major deployments, incidents, and regulatory developments. Together, these findings highlight the need for healthcare AI systems that are not only secure and privacy-preserving, but also transparent, accountable, and trustworthy in real-world clinical settings.

\bibliographystyle{plain}
\bibliography{references,arjun,tushar}

@article{benda2026qualitative,
  title={A qualitative interview study investigating patient, health professional, and developer perspectives on real-world implementation of patient-centered AI systems},
  author={Benda, Natalie and Desai, Pooja and Reza, Zayan and Winogora, Victoria and Suresh, Uday and Zhang, Yiye and Hermann, Alison and Joly, Rochelle and Pathak, Jyotishman and Reading Turchioe, Meghan},
  journal={npj Digital Medicine},
  volume={9},
  number={1},
  pages={352},
  year={2026},
  publisher={Nature Publishing Group UK London}
}

@inproceedings{greshakeNotWhatYouve2023,
	title = {Not what you've signed up for: {Compromising} {Real}-{World} {LLM}-{Integrated} {Applications} with {Indirect} {Prompt} {Injection}},
	shorttitle = {Not what you've signed up for},
	url = {http://arxiv.org/abs/2302.12173},
	doi = {10.48550/arXiv.2302.12173},
	urldate = {2025-12-01},
	publisher = {arXiv},
	author = {Greshake, Kai and Abdelnabi, Sahar and Mishra, Shailesh and Endres, Christoph and Holz, Thorsten and Fritz, Mario},
	month = may,
	year = {2023},
	note = {arXiv:2302.12173 [cs]},
}

@article{collaco2026role,
	title = {The role of agentic artificial intelligence in healthcare: a scoping review},
	journal = {npj Digital Medicine},
	author = {Collaco, Bernardo G and Haider, Syed Ali and Prabha, Srinivasagam and Gomez-Cabello, Cesar A and Genovese, Ariana and Wood, Nadia G and Bagaria, Sanjay P and Gopala, Narayanan and Tao, Cui and Forte, Antonio Jorge},
	year = {2026},
}

@inproceedings{carliniExtractingTrainingData2021,
	title = {Extracting {Training} {Data} from {Large} {Language} {Models}},
	isbn = {978-1-939133-24-3},
	url = {https://www.usenix.org/conference/usenixsecurity21/presentation/carlini-extracting},
	language = {en},
	urldate = {2026-02-18},
	author = {Carlini, Nicholas and Tramèr, Florian and Wallace, Eric and Jagielski, Matthew and Herbert-Voss, Ariel and Lee, Katherine and Roberts, Adam and Brown, Tom and Song, Dawn and Erlingsson, Úlfar and Oprea, Alina and Raffel, Colin},
	year = {2021},
	pages = {2633--2650},
}

@misc{healthArtificialIntelligenceEnabledDeviceMon01/06/2025-09:41,
	title = {Artificial {Intelligence}-{Enabled} {Device} {Software} {Functions}: {Lifecycle} {Management} and {Marketing} {Submission} {Recommendations}},
	shorttitle = {Artificial {Intelligence}-{Enabled} {Device} {Software} {Functions}},
	url = {https://www.fda.gov/regulatory-information/search-fda-guidance-documents/artificial-intelligence-enabled-device-software-functions-lifecycle-management-and-marketing},
	language = {en},
	urldate = {2026-06-11},
	publisher = {FDA},
	author = {Health, Center for Devices {and} Radiological},
	month = jan,
	year = {2025},
}

@article{AIRiskManagement2021,
	title = {{AI} {Risk} {Management} {Framework}},
	url = {https://www.nist.gov/itl/ai-risk-management-framework},
	language = {en},
	urldate = {2026-06-11},
	journal = {NIST},
	month = jul,
	year = {2021},
	note = {Last Modified: 2026-06-10T17:02-04:00},
}

@misc{EthicsGovernanceArtificial,
	title = {Ethics and governance of artificial intelligence for health},
	url = {https://www.who.int/publications/i/item/9789240029200},
	language = {en},
	urldate = {2026-06-11},
}

@article{wysockiAssessingCommunicationGap2023,
	title = {Assessing the communication gap between {AI} models and healthcare professionals: {Explainability}, utility and trust in {AI}-driven clinical decision-making},
	volume = {316},
	issn = {0004-3702},
	shorttitle = {Assessing the communication gap between {AI} models and healthcare professionals},
	url = {https://doi.org/10.1016/j.artint.2022.103839},
	doi = {10.1016/j.artint.2022.103839},
	number = {C},
	urldate = {2026-06-11},
	journal = {Artif. Intell.},
	author = {Wysocki, Oskar and Davies, Jessica Katharine and Vigo, Markel and Armstrong, Anne Caroline and Landers, Dónal and Lee, Rebecca and Freitas, André},
	month = mar,
	year = {2023},
}

@misc{rightsocrHIPAAPrivacyRule2008,
	type = {Page},
	title = {The {HIPAA} {Privacy} {Rule}},
	url = {https://www.hhs.gov/hipaa/for-professionals/privacy/index.html},
	language = {en},
	urldate = {2026-06-11},
	author = {Rights (OCR), Office for Civil},
	month = may,
	year = {2008},
	note = {Last Modified: 2026-03-18},
}

@article{healthArtificialIntelligenceEnabledMedicalTue03/10/2026-09:05,
	title = {Artificial {Intelligence}-{Enabled} {Medical} {Devices}},
	url = {https://www.fda.gov/medical-devices/software-medical-device-samd/artificial-intelligence-enabled-medical-devices},
	language = {en},
	urldate = {2026-06-11},
	journal = {FDA},
	publisher = {FDA},
	author = {Health, Center for Devices {and} Radiological},
	month = mar,
	year = {2026},
}

@misc{EUArtificialIntelligence,
	title = {{EU} {Artificial} {Intelligence} {Act} {\textbar} {Up}-to-date developments and analyses of the {EU} {AI} {Act}},
	url = {https://artificialintelligenceact.eu/},
	language = {en-US},
	urldate = {2026-06-11},
}

@article{mckennaMetaAnalysisVulnerabilityTrust2024,
	title = {A {Meta}-{Analysis} of {Vulnerability} and {Trust} in {Human}–{Robot} {Interaction}},
	volume = {13},
	url = {https://dl.acm.org/doi/10.1145/3658897},
	doi = {10.1145/3658897},
	number = {3},
	urldate = {2026-06-11},
	journal = {ACM Transactions on Human-Robot Interaction},
	author = {Mckenna, Peter E. and Ahmad, Muneeb I. and Maisva, Tafadzwa and Nesset, Birthe and Lohan, Katrin and Hastie, Helen},
	month = aug,
	year = {2024},
	pages = {42:1--42:25},
}

@inproceedings{kaurInterpretingInterpretabilityUnderstanding2020,
	address = {New York, NY, USA},
	series = {{CHI} '20},
	title = {Interpreting {Interpretability}: {Understanding} {Data} {Scientists}' {Use} of {Interpretability} {Tools} for {Machine} {Learning}},
	isbn = {978-1-4503-6708-0},
	shorttitle = {Interpreting {Interpretability}},
	url = {https://dl.acm.org/doi/10.1145/3313831.3376219},
	doi = {10.1145/3313831.3376219},
	urldate = {2026-06-11},
	booktitle = {Proceedings of the 2020 {CHI} {Conference} on {Human} {Factors} in {Computing} {Systems}},
	publisher = {Association for Computing Machinery},
	author = {Kaur, Harmanpreet and Nori, Harsha and Jenkins, Samuel and Caruana, Rich and Wallach, Hanna and Wortman Vaughan, Jennifer},
	month = apr,
	year = {2020},
	pages = {1--14},
}

@inproceedings{bansalDoesWholeExceed2021,
	address = {New York, NY, USA},
	series = {{CHI} '21},
	title = {Does the {Whole} {Exceed} its {Parts}? {The} {Effect} of {AI} {Explanations} on {Complementary} {Team} {Performance}},
	isbn = {978-1-4503-8096-6},
	shorttitle = {Does the {Whole} {Exceed} its {Parts}?},
	url = {https://dl.acm.org/doi/10.1145/3411764.3445717},
	doi = {10.1145/3411764.3445717},
	urldate = {2026-06-11},
	booktitle = {Proceedings of the 2021 {CHI} {Conference} on {Human} {Factors} in {Computing} {Systems}},
	publisher = {Association for Computing Machinery},
	author = {Bansal, Gagan and Wu, Tongshuang and Zhou, Joyce and Fok, Raymond and Nushi, Besmira and Kamar, Ece and Ribeiro, Marco Tulio and Weld, Daniel},
	month = may,
	year = {2021},
	pages = {1--16},
}

@article{kimEstablishingOrganizationalAI2025,
	title = {Establishing organizational {AI} governance in healthcare: a case study in {Canada}},
	volume = {8},
	copyright = {2025 The Author(s)},
	issn = {2398-6352},
	shorttitle = {Establishing organizational {AI} governance in healthcare},
	url = {https://www.nature.com/articles/s41746-025-01909-3},
	doi = {10.1038/s41746-025-01909-3},
	language = {en},
	number = {1},
	urldate = {2026-06-11},
	journal = {npj Digital Medicine},
	publisher = {Nature Publishing Group},
	author = {Kim, Jee Young and Hasan, Alifia and Kueper, Jacqueline and Tang, Terence and Hayes, Chris and Fine, Benjamin and Balu, Suresh and Sendak, Mark},
	month = aug,
	year = {2025},
	pages = {522},
}

@inproceedings{debenedettiAgentDojoDynamicEnvironment2024,
	address = {Red Hook, NY, USA},
	series = {{NIPS} '24},
	title = {{AgentDojo}: a dynamic environment to evaluate prompt injection attacks and defenses for {LLM} agents},
	volume = {37},
	isbn = {979-8-3313-1438-5},
	shorttitle = {{AgentDojo}},
	urldate = {2026-06-11},
	booktitle = {Proceedings of the 38th {International} {Conference} on {Neural} {Information} {Processing} {Systems}},
	publisher = {Curran Associates Inc.},
	author = {Debenedetti, Edoardo and Zhang, Jie and Balunovic, Mislav and Beurer-Kellner, Luca and Fischer, Marc and Tramèr, Florian},
	month = dec,
	year = {2024},
	pages = {82895--82920},
}

@article{ji2023survey,
	title = {Survey of hallucination in natural language generation},
	volume = {55},
	number = {12},
	journal = {ACM computing surveys},
	author = {Ji, Ziwei and Lee, Nayeon and Frieske, Rita and Yu, Tiezheng and Su, Dan and Xu, Yan and Ishii, Etsuko and Bang, Ye Jin and Madotto, Andrea and Fung, Pascale},
	year = {2023},
	pages = {1--38},
}

@article{cutillo2020machine,
	title = {Machine intelligence in healthcare—perspectives on trustworthiness, explainability, usability, and transparency},
	volume = {3},
	number = {1},
	journal = {NPJ digital medicine},
	author = {Cutillo, Christine M and Sharma, Karlie R and Foschini, Luca and Kundu, Shinjini and Mackintosh, Maxine and Mandl, Kenneth D and Noel 1, MI in Healthcare Workshop Working Group Beck Tyler 1 Collier Elaine 1 Colvis Christine 1 Gersing Kenneth 1 Gordon Valery 1 Jensen Roxanne 8 Shabestari Behrouz 9 Southall},
	year = {2020},
	pages = {47},
}

@article{amann2020explainability,
	title = {Explainability for artificial intelligence in healthcare: a multidisciplinary perspective},
	volume = {20},
	number = {1},
	journal = {BMC medical informatics and decision making},
	author = {Amann, Julia and Blasimme, Alessandro and Vayena, Effy and Frey, Dietmar and Madai, Vince I and Consortium, Precise4Q},
	year = {2020},
	pages = {310},
}

@article{tjoa2020survey,
	title = {A survey on explainable artificial intelligence (xai): {Toward} medical xai},
	volume = {32},
	number = {11},
	journal = {IEEE transactions on neural networks and learning systems},
	author = {Tjoa, Erico and Guan, Cuntai},
	year = {2020},
	pages = {4793--4813},
}

@article{el2015anonymising,
	title = {Anonymising and sharing individual patient data},
	volume = {350},
	journal = {bmj},
	author = {El Emam, Khaled and Rodgers, Sam and Malin, Bradley},
	year = {2015},
}

@inproceedings{li2019privacy,
	title = {Privacy-preserving federated brain tumour segmentation},
	booktitle = {International workshop on machine learning in medical imaging},
	publisher = {Springer},
	author = {Li, Wenqi and Milletarì, Fausto and Xu, Daguang and Rieke, Nicola and Hancox, Jonny and Zhu, Wentao and Baust, Maximilian and Cheng, Yan and Ourselin, Sébastien and Cardoso, M Jorge and {others}},
	year = {2019},
	pages = {133--141},
}

@article{kaissis2020secure,
	title = {Secure, privacy-preserving and federated machine learning in medical imaging},
	volume = {2},
	number = {6},
	journal = {Nature Machine Intelligence},
	author = {Kaissis, Georgios A and Makowski, Marcus R and Rückert, Daniel and Braren, Rickmer F},
	year = {2020},
	pages = {305--311},
}

@article{rieke2020future,
	title = {The future of digital health with federated learning. {NPJ} {Digital} {Medicine}, 3, 119},
	journal = {DOI: https://doi. org/10.1038/s41746-020-00323-1},
	author = {Rieke, Nicola and Hancox, Jonny and Li, Wenqi and Milletarì, Fausto and Roth, Holger R and Albarqouni, Shadi and Bakas, Spyridon and Galtier, Mathieu N and Landman, Bennett A and Maier-Hein, Klaus and {others}},
	year = {2020},
}

@article{finlayson2019adversarial,
	title = {Adversarial attacks on medical machine learning},
	volume = {363},
	number = {6433},
	journal = {Science},
	author = {Finlayson, Samuel G and Bowers, John D and Ito, Joichi and Zittrain, Jonathan L and Beam, Andrew L and Kohane, Isaac S},
	year = {2019},
	pages = {1287--1289},
}

@article{heinrichs2025physicians,
	title = {Physicians’ attitudes toward artificial intelligence in medicine: mixed methods survey and interview study},
	volume = {27},
	journal = {Journal of medical Internet research},
	author = {Heinrichs, Helen and Kies, Alexander and Nagel, Saskia K and Kiessling, Fabian},
	year = {2025},
	pages = {e74187},
}

@article{buschMultinationalAttitudesAI2025,
	title = {Multinational attitudes toward {AI} in health care and diagnostics among hospital patients},
	volume = {8},
	number = {6},
	journal = {JAMA network open},
	author = {Busch, Felix and Hoffmann, Lena and Xu, Lina and Zhang, Long Jiang and Hu, Bin and García-Juárez, Ignacio and Toapanta-Yanchapaxi, Liz N and Gorelik, Natalia and Gorelik, Valérie and Rodriguez-Granillo, Gaston A and {others}},
	year = {2025},
	pages = {e2514452},
}

@article{kauttonen2025trust,
	title = {Trust and acceptance challenges in the adoption of {AI} applications in health care: quantitative survey analysis},
	volume = {27},
	journal = {Journal of medical Internet research},
	author = {Kauttonen, Janne and Rousi, Rebekah and Alamäki, Ari},
	year = {2025},
	pages = {e65567},
}

@article{shevtsova2024trust,
	title = {Trust in and acceptance of artificial intelligence applications in medicine: mixed methods study},
	volume = {11},
	number = {1},
	journal = {JMIR human factors},
	author = {Shevtsova, Daria and Ahmed, Anam and Boot, Iris WA and Sanges, Carmen and Hudecek, Michael and Jacobs, John JL and Hort, Simon and Vrijhoef, Hubertus JM},
	year = {2024},
	pages = {e47031},
}

@inproceedings{lee2021included,
	title = {Who is included in human perceptions of {AI}?: {Trust} and perceived fairness around healthcare {AI} and cultural mistrust},
	booktitle = {Proceedings of the 2021 {CHI} conference on human factors in computing systems},
	author = {Lee, Min Kyung and Rich, Katherine},
	year = {2021},
	pages = {1--14},
}

@article{esmaeilzadeh2021patients,
	title = {Patients’ perceptions toward human–artificial intelligence interaction in health care: experimental study},
	volume = {23},
	number = {11},
	journal = {Journal of medical Internet research},
	author = {Esmaeilzadeh, Pouyan and Mirzaei, Tala and Dharanikota, Spurthy},
	year = {2021},
	pages = {e25856},
}

@article{du2024exploring,
	title = {Exploring patient trust in clinical advice from {AI}-driven {LLMs} like {ChatGPT} for self-diagnosis},
	journal = {arXiv preprint arXiv:2402.07920},
	author = {Du, Delong and Paluch, Richard and Stevens, Gunnar and Müller, Claudia},
	year = {2024},
}

@article{kelly2025investigating,
	title = {Investigating {How} {Clinicians} {Form} {Trust} in an {AI}-{Based} {Mental} {Health} {Model}: {Qualitative} {Case} {Study}},
	volume = {12},
	number = {1},
	journal = {JMIR Human Factors},
	author = {Kelly, Anthony and Bhardwaj, Niharika and Sainte-Marie, Trine Theresa Holmberg and Van de Ven, Pepijn and Melia, Ruth and Williams, John Eustis and Mathiasen, Kim and Nielsen, Amalie Søgaard},
	year = {2025},
	pages = {e79658},
}

@article{steerling2025influences,
	title = {Influences on trust in the use of {AI}-based triage—an interview study with primary healthcare professionals and patients in {Sweden}},
	volume = {7},
	journal = {Frontiers in digital health},
	author = {Steerling, Emilie and Svedberg, Petra and Nilsen, Per and Siira, Elin and Nygren, Jens},
	year = {2025},
	pages = {1565080},
}

@article{mesko2023imperative,
	title = {The imperative for regulatory oversight of large language models (or generative {AI}) in healthcare},
	volume = {6},
	number = {1},
	journal = {NPJ digital medicine},
	author = {Meskó, Bertalan and Topol, Eric J},
	year = {2023},
	pages = {120},
}

@article{tu2024towards,
	title = {Towards generalist biomedical {AI}},
	volume = {1},
	number = {3},
	journal = {Nejm Ai},
	author = {Tu, Tao and Azizi, Shekoofeh and Driess, Danny and Schaekermann, Mike and Amin, Mohamed and Chang, Pi-Chuan and Carroll, Andrew and Lau, Charles and Tanno, Ryutaro and Ktena, Ira and {others}},
	year = {2024},
	pages = {AIoa2300138},
}

@article{singhal2023large,
	title = {Large language models encode clinical knowledge},
	volume = {620},
	number = {7972},
	journal = {Nature},
	author = {Singhal, Karan and Azizi, Shekoofeh and Tu, Tao and Mahdavi, S Sara and Wei, Jason and Chung, Hyung Won and Scales, Nathan and Tanwani, Ajay and Cole-Lewis, Heather and Pfohl, Stephen and {others}},
	year = {2023},
	pages = {172--180},
}

@inproceedings{lehman2021does,
	title = {Does {BERT} pretrained on clinical notes reveal sensitive data?},
	booktitle = {Proceedings of the 2021 {Conference} of the {North} {American} {Chapter} of the {Association} for {Computational} {Linguistics}: {Human} {Language} {Technologies}},
	author = {Lehman, Eric and Jain, Sarthak and Pichotta, Karl and Goldberg, Yoav and Wallace, Byron C},
	year = {2021},
	pages = {946--959},
}

@inproceedings{zhang2025agent,
	title = {Agent security bench (asb): {Formalizing} and benchmarking attacks and defenses in llm-based agents},
	volume = {2025},
	booktitle = {International {Conference} on {Learning} {Representations}},
	author = {Zhang, Hanrong and Huang, Jingyuan and Mei, Kai and Yao, Yifei and Wang, Zhenting and Zhan, Chenlu and Wang, Hongwei and Zhang, Yongfeng},
	year = {2025},
	pages = {35331--35366},
}

@misc{willison2023promptinjection,
	title = {Prompt {Injection} {Explained}, with {Video}, {Slides}, and a {Transcript}},
	url = {https://simonwillison.net/2023/May/2/prompt-injection-explained/},
	author = {Willison, Simon},
	month = may,
	year = {2023},
}

@article{liu2023prompt,
	title = {Prompt injection attacks and defenses in llm-integrated applications},
	journal = {arXiv preprint arXiv:2310.12815},
	author = {Liu, Yupei and Jia, Yuqi and Geng, Runpeng and Jia, Jinyuan and Gong, Neil Zhenqiang},
	year = {2023},
}

@article{greshake2023more,
	title = {More than you’ve asked for: {A} comprehensive analysis of novel prompt injection threats to application-integrated large language models},
	volume = {27},
	journal = {arXiv preprint arXiv:2302.12173},
	author = {Greshake, Kai and Abdelnabi, Sahar and Mishra, Shailesh and Endres, Christoph and Holz, Thorsten and Fritz, Mario},
	year = {2023},
}

@misc{commure2026platform,
	title = {Commure: {The} {AI}-{Native} {Enterprise} {RCM} and {Ambient} {Platform}},
	url = {https://www.commure.com},
	author = {{Commure}},
	year = {2026},
}

@misc{patel2025priorauthagents,
	title = {Transform {Healthcare} {Prior} {Authorization} with {AI} {Agents}},
	url = {https://aws.amazon.com/blogs/industries/transform-healthcare-prior-authorization-with-ai-agents/},
	author = {Patel, Manish and Gupta, Vikash and Sundararajan, Laks and Joshi, Ramakant},
	month = nov,
	year = {2025},
}

@misc{microsoft2025dragoncopilot,
	title = {Microsoft {Dragon} {Copilot}},
	url = {https://www.microsoft.com/en-us/health-solutions/clinical-workflow/dragon-copilot},
	author = {{Microsoft}},
	year = {2025},
}

@article{esteva2019guide,
	title = {A guide to deep learning in healthcare},
	volume = {25},
	number = {1},
	journal = {Nature medicine},
	author = {Esteva, Andre and Robicquet, Alexandre and Ramsundar, Bharath and Kuleshov, Volodymyr and DePristo, Mark and Chou, Katherine and Cui, Claire and Corrado, Greg and Thrun, Sebastian and Dean, Jeff},
	year = {2019},
	pages = {24--29},
}

@article{topol2019high,
	title = {High-performance medicine: the convergence of human and artificial intelligence},
	volume = {25},
	number = {1},
	journal = {Nature medicine},
	author = {Topol, Eric J},
	year = {2019},
	pages = {44--56},
}

@article{ma2021understanding,
	title = {Understanding adversarial attacks on deep learning based medical image analysis systems},
	volume = {110},
	journal = {Pattern recognition},
	author = {Ma, Xingjun and Niu, Yuhao and Gu, Lin and Wang, Yisen and Zhao, Yitian and Bailey, James and Lu, Feng},
	year = {2021},
	pages = {107332},
}

@article{fiesler2018reddit,
	title = {Reddit {Rules}! {Characterizing} an {Ecosystem} of {Governance}},
	journal = {ICWSM},
	author = {Fiesler, Casey and {others}},
	year = {2018},
}

@inproceedings{de2014mental,
	title = {Mental health discourse on reddit: {Self}-disclosure, social support, and anonymity},
	booktitle = {{ICWSM}},
	author = {De Choudhury, Munmun and De, Sushovan},
	year = {2014},
}

@inproceedings{aliUnderstandingUsersSecurity2025,
	title = {Understanding {Users}' {Security} and {Privacy} {Concerns} and {Attitudes} {Towards} {Conversational} {AI} {Platforms}},
	isbn = {979-8-3315-2236-0},
	url = {https://www.computer.org/csdl/proceedings-article/sp/2025/223600a298/26EkEQ4fgmA},
	doi = {10.1109/SP61157.2025.00241},
	language = {English},
	urldate = {2026-03-27},
	publisher = {IEEE Computer Society},
	author = {Ali, Mutahar and Arunasalam, Arjun and Farrukh, Habiba},
	month = may,
	year = {2025},
	pages = {298--316},
}

@inproceedings{liItsConsumerBe2023,
	title = {“{It}’s up to the {Consumer} to be {Smart}”: {Understanding} the {Security} and {Privacy} {Attitudes} of {Smart} {Home} {Users} on {Reddit}},
	issn = {2375-1207},
	shorttitle = {“{It}’s up to the {Consumer} to be {Smart}”},
	url = {https://ieeexplore.ieee.org/document/10179344/},
	doi = {10.1109/SP46215.2023.10179344},
	urldate = {2026-03-25},
	booktitle = {2023 {IEEE} {Symposium} on {Security} and {Privacy} ({SP})},
	author = {Li, Jingjie and Sun, Kaiwen and Huff, Brittany Skye and Bierley, Anna Marie and Kim, Younghyun and Schaub, Florian and Fawaz, Kassem},
	month = may,
	year = {2023},
	note = {ISSN: 2375-1207},
	pages = {2850--2866},
}

@misc{caetanoAgenticWorkflowsConversational2025,
	title = {Agentic {Workflows} for {Conversational} {Human}-{AI} {Interaction} {Design}},
	url = {http://arxiv.org/abs/2501.18002},
	doi = {10.48550/arXiv.2501.18002},
	urldate = {2025-12-31},
	publisher = {arXiv},
	author = {Caetano, Arthur and Verma, Kavya and Taheri, Atieh and Kumaran, Radha and Chen, Zichen and Chen, Jiaao and Höllerer, Tobias and Sra, Misha},
	month = jan,
	year = {2025},
	note = {arXiv:2501.18002 [cs]
version: 1},
}

@misc{shimgekarAgenticAIFramework2025,
	title = {Agentic {AI} framework for {End}-to-{End} {Medical} {Data} {Inference}},
	url = {http://arxiv.org/abs/2507.18115},
	doi = {10.48550/arXiv.2507.18115},
	urldate = {2025-12-31},
	publisher = {arXiv},
	author = {Shimgekar, Soorya Ram and Vassef, Shayan and Goyal, Abhay and Kumar, Navin and Saha, Koustuv},
	month = jul,
	year = {2025},
	note = {arXiv:2507.18115 [cs]
version: 1},
}

@misc{llama3,
	title = {{LLaMA} 3: {Open} and efficient foundation models},
	author = {AI, Meta},
	year = {2024},
}

@misc{chatgpt,
  author       = {{OpenAI}},
  title        = {{ChatGPT}},
  year         = {2026},
  url          = {https://chatgpt.com/},
  note         = {Large language model accessed June 9, 2026}
}

@misc{claude,
  author       = {{Anthropic}},
  title        = {{Claude}},
  year         = {2026},
  howpublished = {\url{https://claude.ai/}},
}

@misc{gemini,
  author       = {{Google}},
  title        = {{Gemini}},
  year         = {2026},
  howpublished = {\url{https://gemini.google.com/app}},
  note         = {Google AI assistant. Accessed}
}

@article{singhal2025toward,
  title={Toward expert-level medical question answering with large language models},
  author={Singhal, Karan and Tu, Tao and Gottweis, Juraj and Sayres, Rory and Wulczyn, Ellery and Amin, Mohamed and Hou, Le and Clark, Kevin and Pfohl, Stephen R and Cole-Lewis, Heather and others},
  journal={Nature medicine},
  volume={31},
  number={3},
  pages={943--950},
  year={2025},
  publisher={Nature Publishing Group US New York}
}

@misc{google2025medgemma,
  author       = {{Google Research}},
  title        = {{MedGemma: Our Most Capable Open Models for Health AI Development}},
  year         = {2025},
  month        = jul,
  day          = {9},
  howpublished = {\url{https://research.google/blog/medgemma-our-most-capable-open-models-for-health-ai-development/}},
}

@misc{openai2024gpt4o,
  author       = {{OpenAI}},
  title        = {{Hello GPT-4o}},
  year         = {2024},
  month        = may,
  day          = {13},
  howpublished = {\url{https://openai.com/index/hello-gpt-4o/}},
  note         = {Accessed: 2026-06-09}
}

@article{chen2023meditron,
  title={Meditron-70b: Scaling medical pretraining for large language models},
  author={Chen, Zeming and Cano, Alejandro Hern{\'a}ndez and Romanou, Angelika and Bonnet, Antoine and Matoba, Kyle and Salvi, Francesco and Pagliardini, Matteo and Fan, Simin and K{\"o}pf, Andreas and Mohtashami, Amirkeivan and others},
  journal={arXiv preprint arXiv:2311.16079},
  year={2023}
}

@misc{openai2026healthcare,
  author       = {{OpenAI}},
  title        = {{OpenAI for Healthcare}},
  year         = {2026},
  month        = jan,
  day          = {8},
  howpublished = {\url{https://openai.com/index/openai-for-healthcare/}},
  note         = {Accessed: 2026-06-09}
}

@misc{reddit,
  author       = {{Reddit}},
  title        = {{Reddit}},
  year         = {2026},
  howpublished = {\url{https://www.reddit.com/}},
}

@misc{arcticshift2026,
  author       = {{Arctic Shift}},
  title        = {{Arctic Shift}},
  year         = {2026},
  howpublished = {\url{https://arctic-shift.photon-reddit.com/}},
}

@inproceedings{hurtado2019bot,
  title={Bot detection in reddit political discussion},
  author={Hurtado, Sofia and Ray, Poushali and Marculescu, Radu},
  booktitle={Proceedings of the fourth international workshop on social sensing},
  pages={30--35},
  year={2019}
}

@article{roberta,
  title={Roberta: A robustly optimized bert pretraining approach},
  author={Liu, Yinhan and Ott, Myle and Goyal, Naman and Du, Jingfei and Joshi, Mandar and Chen, Danqi and Levy, Omer and Lewis, Mike and Zettlemoyer, Luke and Stoyanov, Veselin},
  journal={arXiv preprint arXiv:1907.11692},
  year={2019}
}

@inproceedings{bert,
  title={Bert: Pre-training of deep bidirectional transformers for language understanding},
  author={Devlin, Jacob and Chang, Ming-Wei and Lee, Kenton and Toutanova, Kristina},
  booktitle={Proceedings of the 2019 conference of the North American chapter of the association for computational linguistics: human language technologies, volume 1 (long and short papers)},
  pages={4171--4186},
  year={2019}
}

@article{deberta,
  title={Deberta: Decoding-enhanced bert with disentangled attention},
  author={He, Pengcheng and Liu, Xiaodong and Gao, Jianfeng and Chen, Weizhu},
  journal={arXiv preprint arXiv:2006.03654},
  year={2020}
}

@article{phi4,
  title={Phi-4 technical report},
  author={Abdin, Marah and Aneja, Jyoti and Behl, Harkirat and Bubeck, S{\'e}bastien and Eldan, Ronen and Gunasekar, Suriya and Harrison, Michael and Hewett, Russell J and Javaheripi, Mojan and Kauffmann, Piero and others},
  journal={arXiv preprint arXiv:2412.08905},
  year={2024}
}

@article{zhang2023sa,
  title={It’sa fair game, or is it? examining how users navigate disclosure risks and benefits when using llm-based conversational agents},
  author={Zhang, Zhiping and Jia, Michelle and Yao, Bingsheng and Das, Sauvik and Lerner, Ada and Wang, Dakuo and Li, Tianshi},
  journal={arXiv preprint arXiv:2309.11653},
  year={2023}
}

@article{gumusel2024user,
  title={User privacy harms and risks in conversational ai: A proposed framework},
  author={Gumusel, Ece and Zhou, Kyrie Zhixuan and Sanfilippo, Madelyn Rose},
  journal={arXiv preprint arXiv:2402.09716},
  year={2024}
}

@inproceedings{yan2024protecting,
  title={On protecting the data privacy of large language models (llms): A survey},
  author={Yan, Biwei and Li, Kun and Xu, Minghui and Dong, Yueyan and Zhang, Yue and Ren, Zhaochun and Cheng, Xiuzhen},
  booktitle={2024 International Conference on Meta Computing (ICMC)},
  pages={1--12},
  year={2024},
  organization={IEEE}
}

@article{williamson2024balancing,
  title={Balancing privacy and progress: a review of privacy challenges, systemic oversight, and patient perceptions in AI-driven healthcare},
  author={Williamson, Steven M and Prybutok, Victor},
  journal={Applied Sciences},
  volume={14},
  number={2},
  pages={675},
  year={2024},
  publisher={MDPI}
}

@article{proferes2021studying,
  title={Studying reddit: A systematic overview of disciplines, approaches, methods, and ethics},
  author={Proferes, Nicholas and Jones, Naiyan and Gilbert, Sarah and Fiesler, Casey and Zimmer, Michael},
  journal={Social media+ society},
  volume={7},
  number={2},
  pages={20563051211019004},
  year={2021},
  publisher={SAGE Publications Sage UK: London, England}
}

@article{kung2023performance,
  title={Performance of ChatGPT on USMLE: potential for AI-assisted medical education using large language models},
  author={Kung, Tiffany H and Cheatham, Morgan and Medenilla, Arielle and Sillos, Czarina and De Leon, Lorie and Elepa{\~n}o, Camille and Madriaga, Maria and Aggabao, Rimel and Diaz-Candido, Giezel and Maningo, James and others},
  journal={PLoS digital health},
  volume={2},
  number={2},
  pages={e0000198},
  year={2023},
  publisher={Public Library of Science}
}

@article{sellergren2025medgemma,
  title={Medgemma technical report},
  author={Sellergren, Andrew and Kazemzadeh, Sahar and Jaroensri, Tiam and Kiraly, Atilla and Traverse, Madeleine and Kohlberger, Timo and Xu, Shawn and Jamil, Fayaz and Hughes, C{\'\i}an and Lau, Charles and others},
  journal={arXiv preprint arXiv:2507.05201},
  year={2025}
}

@misc{openai2023chatgptoutage,
  author       = {{OpenAI}},
  title        = {March 20 ChatGPT Outage: Here's What Happened},
  year         = {2023},
  month        = mar,
  howpublished = {\url{https://openai.com/index/march-20-chatgpt-outage/}},
  note         = {Published March 24, 2023; accessed June 11, 2026}
}

@online{moveit_2023,
  author  = {{Wikipedia contributors}},
  title   = {2023 {MOVEit} Data Breach},
  year    = {2023},
  url     = {https://en.wikipedia.org/wiki/2023_MOVEit_data_breach},
  note    = {Wikipedia, The Free Encyclopedia}
}

@misc{microsoft_dragon_copilot_2025,
  author       = {{Microsoft Source}},
  title        = {Microsoft {Dragon Copilot} provides the healthcare industry's first unified voice {AI} assistant that enables clinicians to streamline clinical documentation, surface information and automate tasks},
  year         = {2025},
  month        = mar,
  day          = {3},
  url          = {https://news.microsoft.com/source/2025/03/03/microsoft-dragon-copilot-provides-the-healthcare-industrys-first-unified-voice-ai-assistant-that-enables-clinicians-to-streamline-clinical-documentation-surface-information-and-automate-task/},
  note         = {Microsoft News. Accessed: 2026-06-11},
  howpublished = {\url{https://news.microsoft.com/source/2025/03/03/microsoft-dragon-copilot-provides-the-healthcare-industrys-first-unified-voice-ai-assistant-that-enables-clinicians-to-streamline-clinical-documentation-surface-information-and-automate-task/}}
}

@article{madani_chatgpt_salt_2025,
  author       = {Madani, Doha},
  title        = {Man who asked {ChatGPT} about cutting out salt from his diet was hospitalized with hallucinations},
  journal      = {NBC News},
  year         = {2025},
  month        = aug,
  day          = {14},
  url          = {https://www.nbcnews.com/tech/tech-news/man-asked-chatgpt-cutting-salt-diet-was-hospitalized-hallucinations-rcna225055},
  note         = {Accessed: 2026-06-11}
}

@misc{fda_ai_medical_devices,
  author       = {{U.S. Food and Drug Administration}},
  title        = {Artificial Intelligence-Enabled Medical Devices},
  year         = {2026},
  month        = mar,
  url          = {https://www.fda.gov/medical-devices/software-medical-device-samd/artificial-intelligence-enabled-medical-devices},
  note         = {Center for Devices and Radiological Health. Last modified: March 4, 2026. Accessed: 2026-06-11},
  howpublished = {\url{https://www.fda.gov/medical-devices/software-medical-device-samd/artificial-intelligence-enabled-medical-devices}}
}

@misc{eu_ai_act_2024,
  author       = {{European Parliament}},
  title        = {Artificial Intelligence Act},
  year         = {2024},
  month        = mar,
  day          = {13},
  url          = {https://artificialintelligenceact.eu/wp-content/uploads/2024/04/TA-9-2024-0138_EN.pdf},
  note         = {European Parliament legislative resolution on the proposal for a regulation laying down harmonised rules on artificial intelligence (Artificial Intelligence Act). Accessed: 2026-06-11}
}

@misc{whitehouse_2023,
  author       = {{Wikipedia contributors}},
  title        = {Executive Order 14110: Safe, Secure, and Trustworthy Development and Use of Artificial Intelligence},
  year         = {2023},
  url          = {https://en.wikipedia.org/wiki/Executive_Order_14110},
  note         = {Wikipedia, The Free Encyclopedia. Accessed: 2026-06-11},
  howpublished = {\url{https://en.wikipedia.org/wiki/Executive_Order_14110}}
}

@online{hhs_tracking_guidance_2022,
  author  = {{U.S. Department of Health and Human Services}},
  title   = {Use of Online Tracking Technologies by HIPAA Covered Entities and Business Associates},
  year    = {2022},
  url     = {https://www.hhs.gov/hipaa/for-professionals/privacy/guidance/hipaa-online-tracking/index.html},
}

@misc{hr238_119th_2025,
  author       = {{U.S. House of Representatives}},
  title        = {{H.R. 238}: {Healthy Technology Act of 2025}},
  year         = {2025},
  month        = jan,
  day          = {7},
  note         = {119th Congress, 1st Session. Introduced by Rep. David Schweikert (R-AZ). Referred to the House Committee on Energy and Commerce. Accessed: 2026-06-11},
  howpublished = {\url{https://www.congress.gov/bill/119th-congress/house-bill/238}}
}

@article{Abdelwanis,
title = {Artificial intelligence adoption challenges from healthcare providers’ perspectives: A comprehensive review and future directions},
journal = {Safety Science},
volume = {193},
pages = {107028},
year = {2026},
issn = {0925-7535},
doi = {https://doi.org/10.1016/j.ssci.2025.107028},
url = {https://www.sciencedirect.com/science/article/pii/S092575352500253X},
author = {Moustafa Abdelwanis and Mecit Can Emre Simsekler and Adriana F. Gabor and Andrei Sleptchenko and Mohammad Omar},
}

@ARTICLE{fpubh,
AUTHOR={Tang, Zaiyu  and Ma, Wenbao  and Bai, Zhipeng  and Liang, Jiahe  and Xie, Yandong },
TITLE={Public perceptions of AI in healthcare: a large-scale BERTopic and sentiment analysis of Reddit discussions},
JOURNAL={Frontiers in Public Health},
VOLUME={Volume 14 - 2026},
YEAR={2026},
URL={https://www.frontiersin.org/journals/public-health/articles/10.3389/fpubh.2026.1839898},
DOI={10.3389/fpubh.2026.1839898},  
ISSN={2296-2565},  
}

@article{Richardson2021,
  author    = {Richardson, Jordan P. and Smith, Cambray and Curtis, Susan and Watson, Sara and Zhu, Xuan and Barry, Barbara and Sharp, Richard R.},
  title     = {Patient apprehensions about the use of artificial intelligence in healthcare},
  journal   = {npj Digital Medicine},
  year      = {2021},
  issn      = {2398-6352},
  doi       = {10.1038/s41746-021-00509-1},
  url       = {https://doi.org/10.1038/s41746-021-00509-1}
}

@article{rajpurkar2022ai,
  author  = {Rajpurkar, Pranav and Chen, Emma and Banerjee, Oishi and Topol, Eric J.},
  title   = {{AI} in health and medicine},
  journal = {Nature Medicine},
  year    = {2022},
  volume  = {28},
  number  = {1},
  pages   = {31--38},
  doi     = {10.1038/s41591-021-01614-0},
  url     = {https://doi.org/10.1038/s41591-021-01614-0},
  issn    = {1546-170X}
}

@article{moor2023foundation,
  author  = {Moor, Michael and Banerjee, Oishi and Abad, Zahra Shakeri Hossein and Krumholz, Harlan M. and Leskovec, Jure and Topol, Eric J. and Rajpurkar, Pranav},
  title   = {Foundation models for generalist medical artificial intelligence},
  journal = {Nature},
  year    = {2023},
  volume  = {616},
  number  = {7956},
  pages   = {259--265},
  doi     = {10.1038/s41586-023-05881-4},
  url     = {https://doi.org/10.1038/s41586-023-05881-4},
  issn    = {1476-4687}
}

@article{mckinney2020international,
  author  = {McKinney, Scott Mayer and Sieniek, Marcin and Godbole, Varun and Godwin, Jonathan and Antropova, Natasha and Ashrafian, Hutan and Back, Trevor and Chesus, Mary and Corrado, Greg S. and Darzi, Ara and Etemadi, Mozziyar and Garcia-Vicente, Florencia and Gilbert, Fiona J. and Halling-Brown, Mark and Hassabis, Demis and Jansen, Sunny and Karthikesalingam, Alan and Kelly, Christopher J. and King, Dominic and Ledsam, Joseph R. and Melnick, David and Mostofi, Hormuz and Peng, Lily and Reicher, Joshua Jay and Romera-Paredes, Bernardino and Sidebottom, Richard and Suleyman, Mustafa and Tse, Daniel and Young, Kenneth C. and De Fauw, Jeffrey and Shetty, Shravya},
  title   = {International evaluation of an {AI} system for breast cancer screening},
  journal = {Nature},
  year    = {2020},
  volume  = {577},
  number  = {7788},
  pages   = {89--94},
  doi     = {10.1038/s41586-019-1799-6},
  url     = {https://doi.org/10.1038/s41586-019-1799-6},
  issn    = {1476-4687}
}

@article{jiang2023health,
  author  = {Jiang, Lavender Yao and Liu, Xujin Chris and Nejatian, Nima Pour and Nasir-Moin, Mustafa and Wang, Duo and Abidin, Anas and Eaton, Kevin and Riina, Howard Antony and Laufer, Ilya and Punjabi, Paawan and Miceli, Madeline and Kim, Nora C. and Orillac, Cordelia and Schnurman, Zane and Livia, Christopher and Weiss, Hannah and Kurland, David and Neifert, Sean and Dastagirzada, Yosef and Kondziolka, Douglas and Cheung, Alexander T. M. and Yang, Grace and Cao, Ming and Flores, Mona and Costa, Anthony B. and Aphinyanaphongs, Yindalon and Cho, Kyunghyun and Oermann, Eric Karl},
  title   = {Health system-scale language models are all-purpose prediction engines},
  journal = {Nature},
  year    = {2023},
  volume  = {619},
  number  = {7969},
  pages   = {357--362},
  doi     = {10.1038/s41586-023-06160-y},
  url     = {https://doi.org/10.1038/s41586-023-06160-y},
  issn    = {1476-4687}
}

@article{lukac2025ambient,
  author  = {Lukac, P. J. and Turner, W. and Vangala, S. and Chin, A. T. and Khalili, J. and Shih, Y. T. and Sarkisian, C. and Cheng, E. M. and Mafi, J. N.},
  title   = {Ambient {AI} Scribes in Clinical Practice: A Randomized Trial},
  journal = {NEJM AI},
  year    = {2025},
  month   = dec,
  volume  = {2},
  number  = {12},
  doi     = {10.1056/aioa2501000},
  url     = {https://doi.org/10.1056/aioa2501000},
  pmid    = {41497288},
  pmcid   = {PMC12768499}
}

@misc{jahan2025,
      title={Black-Box Behavioral Distillation Breaks Safety Alignment in Medical LLMs}, 
      author={Sohely Jahan and Ruimin Sun},
      year={2025},
      eprint={2512.09403},
      archivePrefix={arXiv},
      primaryClass={cs.LG},
      url={https://arxiv.org/abs/2512.09403}, 
}

\appendix
\appendix

\section{SP\&T Post Corpus Construction}

\subsection{Sub-reddits Related to Healthcare}
\label{subsec:healthcare_subreddits}


Table~\ref{tab:subreddits} lists popular healthcare-related subreddits identified through Reddit keyword searches (e.g., medical AI, healthcare AI, health IT, and AI treatment) and manual exploration. We prioritized active communities with substantial engagement (subscriber counts, weekly contributions) from healthcare professionals, students, and technologists.

\begin{table}[!htbp]
\centering
\caption{Healthcare-related subreddits used in this study.}
\label{tab:subreddits}
\scriptsize
\setlength{\tabcolsep}{2pt}
\renewcommand{\arraystretch}{1.0}
\begin{tabular}{p{0.34\columnwidth}p{0.36\columnwidth}p{0.22\columnwidth}}
\toprule
\textbf{Subreddit} & \textbf{Description} & \textbf{Weekly Eng.} \\
\midrule
\texttt{r/medicine} & Clinical discussion & 244K / 3.3K \\
\texttt{r/healthIT} & EHRs, privacy, security & 23K / 415 \\
\texttt{r/medicalschool} & Medical students & 305K / 5.5K \\
\texttt{r/AskMedical} & Public medical Q\&A & 9.4K / 297 \\
\texttt{r/healthcare} & Systems and policy & 27K / 886 \\
\texttt{r/medschool} & Premed/residents & 82K / 1.8K \\
\texttt{r/Radiology} & Imaging & 138K / 2.7K \\
\texttt{r/askatherapist} & Mental health & 73K / 1.7K \\
\texttt{r/medlabprofessionals} & Lab professionals & 186K / 2.8K \\
\texttt{r/FamilyMedicine} & Family medicine & 82K / 2.1K \\
\texttt{r/HealthInformatics} & Informatics & 4.4K / 93 \\
\texttt{r/HealthTech} & Digital health & 5.9K / 221 \\
\texttt{r/MedicalDevices} & Medical devices & 21K / 541 \\
\texttt{r/nursing} & Nursing & 500K / 18K \\
\bottomrule
\end{tabular}

\vspace{1mm}
{\footnotesize\textit{Note:} Weekly engagement is reported as visitors/posts.}
\end{table}


\subsection{Data Preprocessing}
\label{subsec:data_preprocessing}
We removed Reddit-specific placeholders such as \texttt{[removed]} and \texttt{[deleted]}~\cite{hurtado2019bot}, filtered bot-generated content using known bot identifiers and user-name patterns (e.g., \texttt{AutoModerator}, ``bot''), corrected corrupted Unicode text, normalized whitespace, and deduplicated posts and comments.

\subsection{Keyword-Based Filtering}
\label{subsec:keyword_seed_corpus}

To identify healthcare AI discussions, we constructed a dictionary of 208 AI-related keywords spanning general AI, foundation models, medical AI applications, clinical infrastructure, safety and reliability, regulation, and emerging AI techniques. Keyword matching was performed over each thread's title, selftext, and top five relevant comments. Table~\ref{tab:keyword_categories} summarizes categories  the most frequently matched AI-related terms in the dataset and were used to capture a broad range of healthcare AI discussions. Threads in the collected subreddits contained 9.84 comments on average. Using the top five comments identified 11,981 healthcare AI-related posts, while extending to the top ten comments increased the corpus to 13,591 posts. We therefore use the top five comments as a practical balance between coverage and analytical tractability.


\begin{table}[!htbp]
\centering
\caption{AI keyword categories used for corpus filtering.}
\label{tab:keyword_categories}
\footnotesize
\begin{tabular}{ll}
\toprule
\textbf{Category} & \textbf{Representative Keywords} \\
\midrule
General AI Terms & AI, ML, NLP, Deep Learning \\
Foundation Models & GPT-4, Claude, Gemini, LLaMA \\
Medical AI Applications & Diagnosis, Radiology, Scribe \\
Clinical Infrastructure & EHR, EMR, Telemedicine \\
Safety \& Reliability & Hallucination, Misdiagnosis \\
Regulation \& Compliance & HIPAA, FDA, SaMD \\
Advanced AI Techniques & RAG, Federated Learning, Agents \\
\bottomrule
\end{tabular}
\end{table}


\section{SP\&T Classification Results}
\label{subsec:model_classification_result}

Table~\ref{tab:model_classification_result} presents the performance of models on the SP\&T classification task. Four of the five models achieve accuracy above 0.75, with DeBERTa and RoBERTa performing best (F1-scores of 0.91 and 0.90). Despite variations in performance, the results consistently demonstrate the feasibility of LLM-assisted SP\&T classification across different model architectures.

\begin{table}[!htbp]
\centering
\caption{Model SP\&T classification results.}
\label{tab:model_classification_result}
\footnotesize
\begin{tabular}{lcccc}
\toprule
\textbf{Model} & \textbf{Accuracy} & \textbf{Precision} & \textbf{Recall} & \textbf{F1-Score} \\
\midrule
DeBERTa~\cite{deberta} & 0.92 & 0.87 & 0.95 & 0.91 \\
RoBERTa~\cite{roberta} & 0.91 & 0.85 & 0.96 & 0.90 \\
BERT~\cite{bert}       & 0.76 & 0.68 & 0.69 & 0.69 \\
Llama-3~\cite{llama3}  & 0.75 & 0.72 & 0.62 & 0.67 \\
Phi-4~\cite{phi4}     & 0.56 & 0.47 & 0.69 & 0.55 \\
\bottomrule
\end{tabular}
\end{table}


\section{SP\&T-Related Posts Over Time}
\label{subsec:post-over-time}

\begin{figure}[!htbp]
    \centering
    \includegraphics[width=\linewidth]{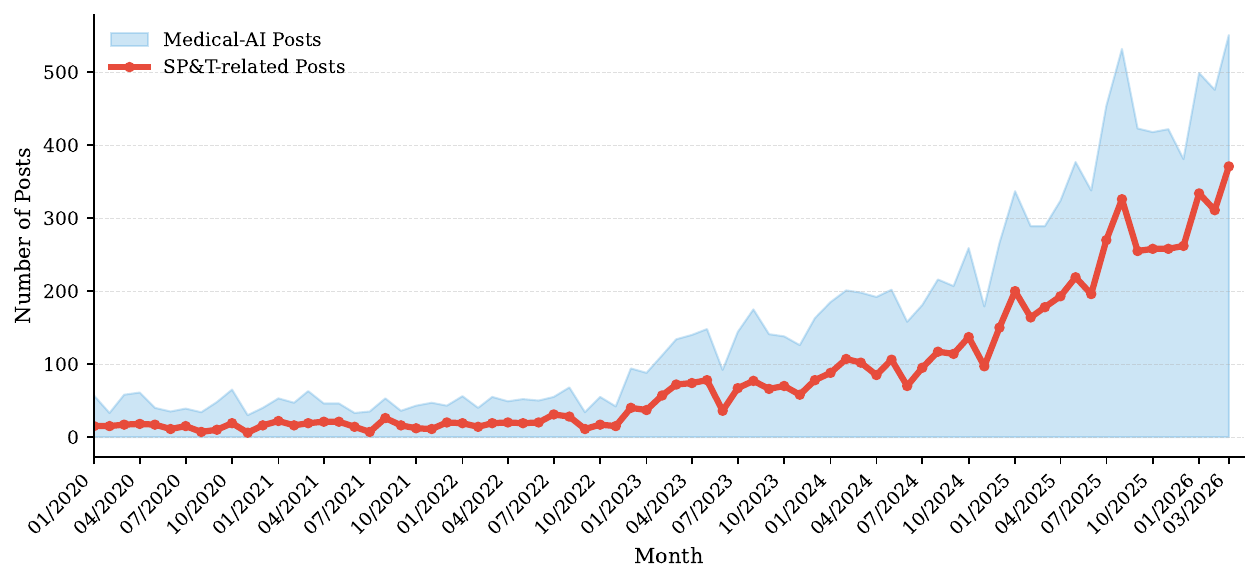}
    \caption{
    Monthly volume of Medical-AI discussions and SP\&T-related discussions from January 2020 to March 2026.
    }
    \label{fig:spt_vs_ai}
\end{figure}

Figure~\ref{fig:spt_vs_ai} shows the volume of SP\&T-related discussions relative to all Medical-AI discussions over time. Of the 11,981 Medical-AI posts identified, 6,514 (54.3\%) discussed at least one SP\&T concern, highlighting the prominence of security, privacy, and trust issues in healthcare AI discussions.

\section{Theme Annotation}
\label{subsec:theme_annotation}

Each SP\&T-related post was assigned a \textit{theme} and \textit{sub-theme} capturing the dominant concern expressed. Table~\ref{tab:codebook} summarizes the six themes and their focus.

\begin{table}[!htbp]
\centering
\caption{Summary of SP\&T concern themes.}
\label{tab:codebook}
\footnotesize
\begin{tabular}{lp{6cm}}
\toprule
\textbf{Theme} & \textbf{Annotation Focus} \\
\midrule
Data Collection & Information collected by healthcare AI systems, including PII, prompts, activity logs, and patient records. \\
Data Usage & How collected data is reused, including model training, research, commercialization, or third-party sharing. \\
Data Retention & Concerns about storage duration, deletion, archiving, or long-term preservation of healthcare data. \\
Security Vulns. & Technical risks such as breaches, unauthorized access, prompt injection, workflow misuse, or AI-agent abuse. \\
Regulatory \& Legal & Compliance concerns involving HIPAA, FDA oversight, EU AI Act, GDPR, or related healthcare regulations. \\
Trust \& Resp. AI & Concerns about safety, bias, explainability, accountability, clinical reliability, and human oversight. \\
\bottomrule
\end{tabular}
\end{table}


\subsection{LLM Annotation Performance}
\label{subsec:llm_annotation_performance}

Table~\ref{tab:theme_subtheme_performance} summarizes the performance of GPT-4o~\cite{openai2024gpt4o} for multi-class classification of user concerns. Overall, GPT-4o achieves strong performance across all concern categories, with F1 scores ranging from 0.88 to 1.00.

\begin{table}[!htbp]
\centering
\caption{Classification performance for SP\&T concerns.}
\label{tab:theme_subtheme_performance}
\footnotesize
\setlength{\tabcolsep}{6pt}
\begin{tabular}{@{}lccc@{}}
\toprule
\textbf{Type of Concern} & \textbf{Prec.} & \textbf{Rec.} & \textbf{F1} \\
\midrule
\textbf{Data Collection} & \textbf{0.95} & \textbf{0.86} & \textbf{0.90} \\
\hspace{0.3cm}PII & 0.91 & 0.91 & 0.91 \\
\hspace{0.3cm}Activity/Behavioral Data & 0.90 & 0.90 & 0.90 \\
\midrule
\textbf{Data Usage} & \textbf{1.00} & \textbf{0.83} & \textbf{0.91} \\
\hspace{0.3cm}Model Training & 1.00 & 1.00 & 1.00 \\
\hspace{0.3cm}Third-Party Sharing & 1.00 & 0.67 & 0.80 \\
\midrule
\textbf{Data Retention} & \textbf{1.00} & \textbf{1.00} & \textbf{1.00} \\
\midrule
\textbf{Security Vulnerabilities} & \textbf{1.00} & \textbf{0.92} & \textbf{0.96} \\
\hspace{0.3cm}Platform Security & 1.00 & 1.00 & 1.00 \\
\hspace{0.3cm}AI Workflow Security & 0.78 & 0.88 & 0.82 \\
\midrule
\textbf{Regulatory \& Legal Compliance} & \textbf{0.90} & \textbf{0.86} & \textbf{0.88} \\
\hspace{0.3cm}HIPAA & 1.00 & 0.88 & 0.93 \\
\hspace{0.3cm}FDA & 0.60 & 1.00 & 0.75 \\
\hspace{0.3cm}EU Regulation & 1.00 & 1.00 & 1.00 \\
\midrule
\textbf{Trust \& Responsible AI} & \textbf{0.95} & \textbf{0.99} & \textbf{0.97} \\
\hspace{0.3cm}Safety & 0.95 & 0.97 & 0.96 \\
\hspace{0.3cm}Bias & 1.00 & 1.00 & 1.00 \\
\hspace{0.3cm}Explainability & 1.00 & 1.00 & 1.00 \\
\hspace{0.3cm}Human Oversight & 0.87 & 0.81 & 0.84 \\

\bottomrule
\end{tabular}
\end{table}

\subsection{Medical AI Events for Time-Series Analysis}
Table~\ref{tab:event_category_results} lists the events and their dates and
reports estimated immediate changes and shifts in weekly discussion trends
for each event--category pair, with $p$-value of 0.05.

\begin{table}[!htbp]
\centering
\caption{List of key Medical AI events and changes in the level and trend of weekly SP\&T discussions following selected events, by concern category.}
\label{tab:event_category_results}
\scriptsize
\setlength{\tabcolsep}{2.5pt}
\begin{tabular}{@{}llrrrrrl@{}}
\toprule
Event & Date & $\beta_I$ & $p_I$ & $\beta_T$ & $p_T$ & $R^2$ & Sig.\\

\midrule
\multicolumn{8}{l}{\textbf{AI Development}} \\
\midrule
ChatGPT & 2022-11-30 & 26.22 & \textbf{0.0006} & -1.21 & 0.1016 & 0.807 & $I^+$ \\
Med-PaLM & 2022-12-31 & 29.93 & \textbf{0.0005} & -4.69 & \textbf{0.0003} & 0.746 & $I^+,T^-$ \\
GPT-4 & 2023-03-14 & 2.68 & 0.8922 & 1.76 & 0.4524 & 0.279 & -- \\
Med-PaLM 2 & 2023-03-31 & 54.60 & \textbf{0.0021} & -3.49 & 0.0576 & 0.620 & $I^+$ \\
Claude 2 & 2023-07-11 & 3.32 & 0.4668 & 0.14 & 0.7917 & 0.189 & -- \\
MEDITRON-70B & 2023-11-30 & -2.73 & 0.5178 & 0.46 & 0.3567 & 0.073 & -- \\
Gemini 1.0 & 2023-12-06 & -1.52 & 0.6779 & 0.61 & 0.1701 & 0.240 & -- \\
GPT-4o & 2024-05-13 & -7.49 & 0.2143 & 0.53 & 0.4455 & 0.257 & -- \\
Claude 3.5 Sonnet & 2024-06-21 & 8.93 & 0.2689 & 0.04 & 0.9672 & 0.552 & -- \\
Gemini 2.0 & 2024-12-11 & 10.86 & 0.0896 & -1.05 & 0.1535 & 0.206 & -- \\
Data Poisoning Attacks & 2024-12-31 & 6.41 & 0.2589 & -0.74 & 0.2666 & 0.149 & -- \\
MedGemma & 2025-05-30 & 35.04 & \textbf{0.0048} & -2.03 & 0.1177 & 0.551 & $I^+$ \\
\midrule
\multicolumn{8}{l}{\textbf{Data Privacy}} \\
\midrule
ChatGPT Data Leak & 2023-03-31 & 19.27 & \textbf{0.0066} & -1.38 & 0.0704 & 0.487 & $I^+$ \\
MOVEit Breach & 2023-05-31 & 28.17 & \textbf{0.0096} & -0.81 & 0.4714 & 0.611 & $I^+$ \\
Personal Health LLM & 2024-06-30 & 1.07 & 0.5222 & -0.16 & 0.4271 & 0.185 & -- \\
Dragon Copilot & 2025-03-03 & 2.89 & 0.5816 & -0.45 & 0.4729 & 0.074 & -- \\
Episource Breach & 2025-06-16 & 4.54 & 0.5442 & -0.66 & 0.4574 & 0.047 & -- \\

\midrule
\multicolumn{8}{l}{\textbf{Human Oversight and Explainability}} \\
\midrule
Amazon Care Shutdown & 2022-08-22 & 1.93 & 0.5022 & -0.07 & 0.8367 & 0.083 & -- \\
Ban on AI Prescriptions & 2023-03-15 & 1.59 & 0.7810 & -0.33 & 0.6247 & 0.063 & -- \\
Dragon Copilot & 2025-03-03 & 47.52 & \textbf{0.0001} & -3.71 & \textbf{0.0031} & 0.711 & $I^+,T^-$ \\
1,250+ AI Devices & 2025-07-31 & 53.99 & \textbf{0.0120} & -4.15 & 0.0787 & 0.520 & $I^+$ \\
ChatGPT Misdiagnosis & 2025-09-15 & 32.67 & 0.0785 & -3.87 & 0.0766 & 0.484 & -- \\

\midrule
\multicolumn{8}{l}{\textbf{Regulatory and Security Governance}} \\
\midrule
HIPAA Tracking Guidance & 2022-12-31 & 9.79 & 0.0569 & -0.37 & 0.5123 & 0.507 & -- \\
White House Order on AI & 2023-10-31 & 16.66 & \textbf{0.0010} & -1.16 & \textbf{0.0275} & 0.613 & $I^+,T^-$ \\
EU AI Act & 2024-03-31 & 18.23 & \textbf{0.0392} & -1.18 & 0.2277 & 0.338 & $I^+$ \\
Healthy Tech Act 2025 & 2025-02-05 & 14.78 & \textbf{0.0037} & -1.21 & \textbf{0.0292} & 0.504 & $I^+,T^-$ \\

\bottomrule
\end{tabular}
\vspace{2pt}

\parbox{0.5\textwidth}{\scriptsize
\textit{Note.}
$\beta_I$ and $\beta_T$ are the intervention and time--intervention
coefficients, respectively; each model uses 16 weekly observations. Positive denotes at least one significant positive coefficient, even when another coefficient is significantly
negative.$I^+$ and $I^-$ indicate significant positive and
negative immediate changes, respectively;
$T^+$ and $T^-$ indicate significant positive and
negative trend changes.}
\end{table}

\end{document}